\documentclass[10pt,twocolumn,A4]{article}

\usepackage{amsmath}
\usepackage{amssymb}
\usepackage{graphicx}
\usepackage{cite}
\usepackage[small,bf]{caption}

\usepackage{xspace}

\newcommand{\Eq}[1]{Eq.\,(\ref{#1})}
\newcommand{\Fig}[1]{Fig.\,\ref{#1}}
\newcommand{\Sec}[1]{Sec.\,\ref{#1}}

\newcommand{\ie}{i.\,e.\xspace}
\newcommand{\eg}{e.\,g.\xspace}
\newcommand{\LCDM}{$\Lambda$CDM\xspace}

\def\kms{\ifmmode{{\rm km \, s^{-1}}}\else{${\rm km \, s^{-1}}$}\xspace\fi}
\def\Mchar{\ifmmode{{M_{1/2}}}\else{$M_{1/2}$}\xspace\fi}
\def\vchar{\ifmmode{{v_{1/2}}}\else{$v_{1/2}$}\xspace\fi}
\def\vmax{\ifmmode{{v_{\rm max}}}\else{$v_{\rm max}$}\xspace\fi}
\def\vcirc{\ifmmode{{v_{\rm circ}}}\else{$v_{\rm circ}$}\xspace\fi}

\def\hMpc{\ifmmode{\:h^{-1}\,{\rm Mpc}}\else{$h^{-1}\,{\rm Mpc}$}\fi}
\def\hkpc{\ifmmode{h^{-1} \,{\rm kpc}}\else{$h^{-1}\,{\rm kpc}$}\fi}
\def\hMsun{\ifmmode{\:h^{-1}\,M_\odot}\else{$h^{-1}\,M_\odot$}\fi}

\begin{document}

\title{Dwarf Galaxies in Voids: \\
        Dark Matter Halos and Gas Cooling}
\author{Matthias Hoeft$^1$ \& Stefan Gottl\"ober$^2$ \\
        \small 
        $^1$Th\"uringer Landessternwarte Tautenburg, 07778 Tautenburg, Germany \\ 
        \small
        $^2$Astrophysikalisches Institut Potsdam, 14482 Potsdam, Germany  }
\maketitle
\date{}

\abstract{
  Galaxy surveys have shown that luminous galaxies are mainly distributed in
  large filaments and galaxy clusters.  The remaining large volumes are
  virtually devoid of luminous galaxies.  This is in concordance with the
  formation of the large-scale structure in Universe as derived from
  cosmological simulations. However, the numerical results indicate that
  cosmological voids are abundantly populated with dark matter haloes which
  may in principle host dwarf galaxies.  Observational efforts have in
  contrast revealed, that voids are apparently devoid of dwarf galaxies. We
  investigate the formation of dwarf galaxies in voids by hydrodynamical
  cosmological simulations.  Due to the cosmic ultra-violet background
  radiation low-mass haloes show generally are reduced baryon fraction.  We
  determine the characteristic mass below which dwarf galaxies are baryon
  deficient. We show that the circular velocity below which the accretion of
  baryons is suppressed is approximately $40 \,\kms$.  The suppressed baryon
  accretion is caused by the photo-heating due to the UV-background.  We set up
  a spherical halo model and show that the effective equation of state of the
  gas in the periphery of dwarf galaxies determines the characteristic mass.
  This implies that any process which heats the gas around dwarf galaxies
  increases the characteristic mass and thus reduces the number of observable
  dwarf galaxies.  
  }

\vfill
\noindent
{\small\it send offprint requests to: hoeft@tls-tautenburg.de }


\section{Introduction: Formation of Structure in the Universe}

  During the last couple of decades new extensive observations of the Universe
  were made using both ground-based telescopes and space instruments. These
  measurements have provided new insights into the structure of the Universe
  on various scales. A wide range of the electromagnetic spectrum emitted by
  cosmic objects has been studied. The wavelengths extend from very long radio
  wavelengths to energetic gamma rays. This observational progress has been
  accompanied by considerable effort in our theoretical understanding of the
  formation of different components of the observed structure of the Universe
  from small scales (galaxies and their satellites) up to the largest scales
  (clusters of galaxies and superclusters).
  \\

  The standard picture of structure formation was suggested by White and Rees
  \cite{WR78} more than 30 years ago. It suggests that gravitational
  instability drives the dark matter (DM) to cluster hierarchically in a
  bottom-up fashion. The gaseous baryons settle into to the DM haloes (namely,
  bound virial DM structures) and via gas-dynamical dissipative processes
  cool, fragment and form stars. This simple picture prevails today, although
  our knowledge of the details and of the actual processes has evolved
  dramatically since then. The non-linear nature of the gravitational dynamics
  and the gas-astrophysical processes makes the problem of structure formation
  virtually intractable analytically, and therefore the fiecp ..ld relies on
  numerical simulations. A substantial part of the progress in understanding
  structure formation in the universe is due to the increasing possibilities
  to make numerical experiments using the largest massive parallel
  supercomputers. In the eighties the best simulations handled $32^3$
  particles whereas at present $1024^3$ particles became a standard for
  numerical simulations, a increase of $2^{15}$ roughly in agreement with
  Moore's law.
  \\

  The effort of observers and theorists brought about the so called standard
  cosmological model. This model is based on the idea that some kind of dark
  energy contributes about 70\,\% of the total energy density of the spatially
  flat Universe in which the total energy density equals the critical one. The
  dark energy is responsible for the observed accelerated expansion of the
  universe. The simplest form of the dark energy is the cosmological constant,
  which was introduced already in 1917 by Albert Einstein in his paper about
  the cosmological solutions of the field equations of general relativity. The
  remaining about 30\,\% of energy density consists of matter. About 85\,\% of
  this matter is made of unknown dark matter particles, the remaining 15\,\%
  is the contribution of ``normal'' baryonic particles well known to particle
  physicists. This means that the nature of more than 95\,\% of the matter in
  the Universe is not yet understood.
  \\

  According to this standard cosmological model, the main process responsible
  for the formation of the observed structures is gravitational instability.
  The initial seeds, which eventually became galaxies and superclusters and
  all other structures, came from the quantum fluctuations generated during
  the early inflationary phase: $\sim 10^{-35}$\,sec or so from the beginning
  of the Big Bang. The power spectrum of these primordial fluctuations has
  been confirmed by measuring the temperature fluctuations of the cosmic
  microwave background radiation. These temperature fluctuations tell us the
  magnitude of the small fluctuations in the Universe about 300\,000 years
  after the Big Bang.
  \\

  One of the key features of the standard model is its simplicity. The
  expansion rate and the clustering properties are described by only few
  parameters which are measured at present with quite high accuracy. These
  parameters are the current rate of universal expansion, $H_0 = h \times
  100\, {\rm km \, s^{-1} \, Mpc^{-1}}$, the mass density parameter,
  $\Omega_{\rm mat}$, the value of the cosmological constant,
  $\Omega_{\Lambda}$, the primordial baryon abundance, $\Omega_{\rm b}$, and
  the overall normalisation of the power spectrum of initial density
  fluctuations, typically characterised by $\sigma_8$, the present-day r.m.s.
  mass fluctuations on spheres of radius $8\,h^{-1}$ Mpc.
  \\

  The initial power spectrum (created during the inflationary phase) does not
  contain preferred length scales. However, the horizon size at
  matter-radiation equality is mapped on this scale free spectrum. Since this
  scale (today around 100 Mpc) is well above the typical scales of observed
  objects structure formation is predicted to be essentially scale invariant:
  in a statistical sense the structures on scales of galaxy clusters are
  repeated on scales of galaxies. Typically, small objects merge together and
  form more and more massive objects. However, the small objects do not
  disappear within those larger objects but rather form a complex hierarchy of
  substructures. This hierarchical scenario predicts that our Milky Way
  galaxy is expected to have as many satellites (many hundreds) as a cluster
  of galaxies has galaxies. However, only a few dozen satellite of the Milky
  Way have been observed yet: this is the well known missing satellite problem
  \cite{Klypinetal1999, Mooreetal1999}. Other manifestations of the scale free
  power spectrum are the predicted large number of dwarfs in low density
  regions of the universe \cite{Peebles2001,gottloeber:03} or the predicted
  spectrum of mini-voids in the Local Universe \cite{TK08}.
  \\

  A better understanding of the physics of structure formation on small
  scales, in particular of the correct modelling of baryonic physics, could
  solve these problems \cite{hoeft:06}. Based on semi-analytical models
  recently Maccio {\it et al.} \cite{Maccio2009} (see also Ref.
  \cite{Koposov2009}) claimed that the long standing problem of missing
  satellites can be solved within the $\Lambda$CDM scenario. The basic idea
  behind this solution of the problem is that the haloes which host a galaxy
  of a given measured rotational velocity are more massive than expected by
  the direct association of rotational velocity and the haloes maximum
  circular velocity. Since more massive haloes are less frequent the problem
  of missing satellites is solved. However, any non-baryonic physics that
  reduces power on small scales compared to the standard model will also
  improve the situation. It is well known that warm dark matter acts in this
  direction by erasing power at short scales due to free streaming
  \cite{Tikhonovetal2009,Zavalaetal2009}.
  \\

  Altogether, we arrive at a picture in which dark matter particles form the
  backbone structure for all objects in the Universe from clusters of galaxies
  to dwarf galaxies. Normal matter (baryons) falls into the potential wells
  formed by the dark matter particles and forms the luminous objects. The
  details of this formation process must be followed using hydrodynamical
  simulations. Galaxy formation depends on many physical processes, some of
  which are very poorly understood, and some of which take place on sub-grid
  scales and therefore cannot be modelled fully and consistently and need to
  be fudged numerically by ad-hoc recipes. For example the formation of stars
  and the feedback of stars on the intergalactic medium take place on scales
  orders of magnitudes below the present day resolution. In cosmological
  simulations of galaxy formation typical masses of ``stellar particles'' are
  $10^4 \hMsun$ or more. Thus such a particle represents the formation and
  evolution of a large number of real stars.
  \\

  The paper is organised as follows: In Sec\:2 we introduce briefly the method
  of cosmological simulations and discuss observational facts about voids and
  how they are defined in simulations. In Sec.\:3 we present the baryon
  fraction in low mass dark matter haloes and introduce the characteristic
  mass. In Sec.\:4 we introduce a spherical model for the halo gas and discuss
  the relation between photo-heating of the halo gas and gas accretion. Our
  findings are summarised in Sec.\:5.

  \vfill
  

\section{Cosmological simulations}

  The cosmic microwave background (CMB) originates from the recombination of
  the hydrogen in the Universe at redshift about 1000. Primordial small
  density fluctuations are imprinted on the spectrum of the CMB temperature
  fluctuations which is closely related to the initial power spectrum of
  density fluctuations. The observed unique features of CMB temperature
  fluctuations constrain the cosmological parameters as well as the
  normalisation of the power spectrum of density fluctuations. The most recent
  results of the WMAP experiment combined with the measurements of the
  baryonic oscillations and supernova data \cite{Hinshawetal2008} yield a
  Hubble parameter $h = 7.01$, the density of dark matter $\Omega_{\rm DM} =
  0.228$, the baryonic density $\Omega_{\rm bar} = 0.046$, the cosmological
  constant $\Omega_{\Lambda} = 0.726$, the normalisation of the power spectrum
  $\sigma_8 = 0.82$ and its slope $n=0.96$. These parameters fix the
  cosmological model and therefore also the power spectrum of the initial
  perturbations. Starting with the initial conditions the simulations follow
  the growth of the perturbation in a universe that expands according to the
  chosen cosmological parameters. In this expanding universe the gravitational
  interaction is Newtonian.

\subsection{Initial conditions}

  The first step of running cosmological simulations is to set up the initial
  conditions, namely amplitudes and phases of small perturbations at a very
  high redshift. Having in mind that the largest structures in the universe --
  superclusters and voids -- have sizes of 10 to 50\,Mpc, the simulated volume
  should be significantly larger. However, we may be interested in the
  structure of a much smaller object such as our Milky Way galaxy or its
  satellites. In $N$-body simulations each mass element is represented by a
  point-like particle. The mass resolution is limited by the total number of
  particles computers can handle at a given time. Thus, increasing the
  representative cosmological volume decreases the mass resolution. To
  overcome this problem mass refinement techniques have been developed. Here
  we follow the algorithm proposed in Ref.\,\cite{klypin:01}: To construct
  suitable initial conditions, we first create a random realisation at the
  highest possible resolution. This depends on the available computers; at
  present we reach $4096^3$ particles. The initial displacements and
  velocities of $N$ particles are calculated using all waves ranging from the
  fundamental mode $k=2\pi/L_{\rm box}$ to the Nyquist frequency $k_{\rm
  Ny}=2\pi/L_{\rm box}\times N^{1/3}/2$. Then the resolution can be decreased
  by replacing $8^i$ ($i = 1,5$) neighbouring particles by one particle with
  $8^i$ higher mass which results in a distribution of $2048^3$, $1024^3$,
  $512^3$, $256^3$ or $128^3$ particles. Using a smaller number of more
  massive particles we first run low-resolution simulations until the present
  epoch. In this simulation we select the regions of interest. Then we repeat
  the simulation but this time we preserve low mass particles inside the
  region of interest. Outside of this region we progressively replace small
  particles by massive ones creating shells of more and more massive particles
  until we reach the low resolution region of $128^3$ particles. This
  procedure ensures that our object evolves in the proper cosmological
  environment and with the right gravitational tidal fields.
  \\

  Based on the power spectrum of density fluctuations we set up the initial
  conditions of our simulation at an early redshift which depends on the size
  of the box and the required resolution. To move the particles from the
  original Lagrangian point on a regular grid to their Eulerian position we
  use the Zeldovich approximation. The power spectrum of the generated
  Gaussian stochastic density field corresponds to the input power spectrum.
  The random nature of the realisation manifests itself on scales comparable
  to the box size where the spectrum is sampled only by a few modes. By this
  procedure the cosmological parameters and the normalisation of the density
  power spectrum fully determine the initial conditions of the cosmological
  simulations, except of a random number which characterises the starting
  point of the random number generator. Thus, on scales comparable to the box
  size cosmic variance enters through this number which characterises a given
  simulation. On smaller scales - typically a quarter of the box size - the
  realisation follows very closely the input power spectrum. Starting from the
  initial matter distribution we simulate the formation of cosmological
  structures. More precisely, we integrate the Poisson equation by the help of
  particle methods (Gadget and ART).

\subsection {The simulation  method}

  Our hydrodynamical simulations have been run with an updated version of the
  parallel Tree-SPH code Gadget \cite{springel:01,springel:02}. The dark and
  baryonic matter distributions are represented by particles. The
  gravitational forces are computed by a new algorithm based on the Tree-PM
  method which speeds up the force computation significantly compared with a
  pure tree algorithm. The hydrodynamical equations are solved by a
  Smoothed-Particle Hydrodynamics (SPH) method. The code uses an
  entropy-conserving formulation of SPH \cite{springel:02} which alleviates
  problems due to numerical over-cooling.
  \\

  The code also includes photo-ionisation and radiative cooling processes for
  an optically thin primordial mix of helium and hydrogen. For computing the
  thermal evolution it uses Êionisation, recombination and cooling rates as
  given in Ref.\,\cite{katz:96}. Since we will set up later in this paper an
  analytic model for the evolution of the intergalactic medium (IGM) we give
  here in detail the processes which affect the temperature, $T$, of the IGM.
  It primarily changes due to photo-heating, ${\cal H}$, and radiative
  cooling, $\Gamma$. Moreover, it decreases due to the adiabatic Hubble
  expansion, ${\rm d}T \propto H\,{\rm d}t$, where $H$ is the time-dependent
  Hubble constant. Cosmological structure formation changes the temperature
  adiabatically, ${\rm d}T \propto {\rm d}\rho$. Finally, changes of the
  chemical composition also affect the temperature, ${\rm d}T \propto \,{\rm
  d}\Sigma_i X_i$, where $X_i$ denotes the abundance of the atomic species $i$
  with respect to the cosmic baryon density, $X_i = n_i/n_{\rm b}$. In
  summary, the temperature evolution is given by 
  \begin{equation} 
  \begin{split}
    {\rm d}T
    = \:
      \frac{2}{3} 
      \frac{({\cal H} - \Gamma) \, {\rm d}t}
           {k_{\rm B} n _{\rm b} \Sigma_i X_i}
    - \: 2 H T \, {\rm d}t
    \\
    +  \: \frac{2}{3} 
      \frac{T \, {\rm d} \Delta }
           { \Delta }
    - \: \frac{T \, {\rm d}(\Sigma_i X_i)}
           {\Sigma_i X_i} 
    ,
    \end{split}
    \label{eq-temp-evolve}
  \end{equation}
  where $\Delta$ denotes the overdensity, $\Delta = \rho/\langle\rho\rangle$.
  We take the atomic hydrogen and helium species, \{HI, HII, HeI, HeII,
  HeIII\}, into account. We assume a constant, primordial helium mass fraction
  of $Y_{\rm p}=0.24$.  
  \\

  The physics of star formation is treated in the code by means of a
  sub-resolution model in which the gas of the interstellar medium (ISM) is
  described as a multiphase medium of hot and cold gas \cite{yepes:97,
  springel:03}. Cold gas clouds are generated due to cooling and they are the
  material out of which stars can be formed in regions that are sufficiently
  dense. Supernova feedback heats the hot phase of the ISM and evaporates cold
  clouds, thereby establishing a self-regulation cycle for star formation. The
  heat input due to the supernovae also leads to a net pressurisation of the
  ISM, such that its effective equation of state becomes stiffer than
  isothermal. This stabilises the dense star forming gas in galaxies against
  further gravitational collapse, and allows converged numerical results for
  star formation even at moderate resolution. See \cite{springel:03} for a
  more detailed description of the star formation model implemented in the
  Gadget code.

\subsection{Dark matter haloes}

  During the cosmological evolution dark matter haloes are formed which
  accrete more and more matter or merge with other ones. Within a simulation
  with one billion particles it is a challenge to find structures and
  substructures.  To find structures at virial over-density one can use a
  friends-of-friends algorithm with a certain linking length (0.17 of the mean
  inter-particle distance for the \LCDM model at redshift zero). The resulting
  particle clusters are in general tri-axial objects. Substructures can be
  identified as particle clusters at smaller linking lengths (higher
  over-densities). The more different linking lengths are used the better
  substructures will be resolved. Thus a whole hierarchy of friends-of-friends
  clusters have to be calculated. To this end we have developed a hierarchical
  friends-of-friends algorithm which is based on the calculation of the
  minimum spanning tree of the given particle distribution. We use also the
  bound-density-maxima (BDM) algorithm \cite{Klypin1999} which determines
  spherical haloes and their sub-haloes. The code removes unbound particles
  which is particularly important for sub-haloes.
  \\

  At all redshifts the sample of haloes is characterised by the mass function
  of the isolated haloes. This number density of haloes of a given mass
  depends on the power spectrum. A first very successful analytical ansatz to
  predict the mass function of haloes has been made by Press and Schechter
  \cite{PS1974}. Later on this has been improved by Bond {\it et al.}
  \cite{Bondetal1991} and Sheth and Tormen \cite{ST1999}. The search for an
  accurate and universal function which describes the number density of haloes
  found in simulation at different redshifts led to a set of fitting functions
  proposed by different authors \cite{Jenkinsetal2001, Warrenetal2006,
  Tinkeretal2008}.

\subsection{Voids}

  With the first available large galaxy surveys it became clear that there
  exist large regions in the Universe which are not occupied by bright
  galaxies \cite{gregory:78, Joeveer1978, Kirshner81}. Regions of all possible
  sizes devoid of galaxies can be seen in all redshift surveys. The
  observational discovery was soon followed by the theoretical understanding
  that voids constitute a natural outcome of structure formation via
  gravitational instability \cite{Peebles1982, Hoffman1982}. Together with
  clusters, filaments, and superclusters, giant voids constitute the
  large-scale structure of the Universe.
  \\

  More than 20 years ago, for the first time the sizes of voids have been
  estimated in different samples of galaxies with measured redshifts
  \cite{Einasto1989}. Later on voids in the CfA redshift catalogues were
  studied by \cite{Vogeley1994}, in the Las Campanas redshift survey by
  \cite{MuellerLCRS}, in the Optical Redshift Survey and in the IRAS 1.2-Jy
  survey by \cite{ElAd1997, ElAd2000} and in the PSCz catalogue by Plionis and
  Basilakos \cite{Plionis2002} and Hoyle and Vogeley \cite{HoyleVogeley}. For
  a review of early observational efforts see Peebles \cite{Peebles2001}. More
  recently voids have been studied by many authors using the Two-Degree Field
  Galaxy Redshift Survey \cite{HoyleVogeley2004, Patirietal2006,
  Ceccarellietal2006, Crotonetal2004} and the Sloan Digital Sky Survey
  \cite{Rojasetal2004, Vogeleyetal2004, goldberg:05, Hoyleetal2005}.
  \\

  There were several attempts to find dwarf galaxies in few individual voids
  \cite{Lindner1996, Popescu1997, Kuhn1997, grogin:99}. The overall conclusion
  is that faint galaxies do not show a strong tendency to fill up voids
  defined by bright galaxies. The strongest arguments that voids are not
  populated by dwarf galaxies were given by Peebles \cite{Peebles2001} who
  points out that the dwarf galaxies in the ORS catalogue follow remarkably
  close the distribution of bright galaxies: there are no indications that
  they fill voids in the distribution of bright galaxies. To summarise,
  observations indicate that large voids found in the distribution of bright
  $(\sim M_*)$ galaxies are empty of galaxies, which are two magnitudes below
  $M_*$. Recently, Tikhonov and Klypin \cite{TK08} studied the properties of
  mini-voids in the nearby universe.
  \\

  The void phenomenon was already the target of many theoretical studies
  \cite{Einasto1991, Sahni1994, Ghigna1994, Ghigna1996, Friedmann2001,
  Mathis2002, Bensonetal2002, Shandarinetal2006, Furlanetto2006, LeePark2006}.
  Using a set of DM only simulations \cite{gottloeber:03} have studied the
  inner structure of voids. The haloes in voids are arranged in a pattern,
  which looks like a miniature Universe with the same structural elements as
  the large-scale structure of the galactic distribution of the Universe.
  There are filaments and voids; larger haloes are at the intersections of
  filaments. The mass function of haloes in voids is much steeper for high
  masses resulting in very few galaxies with circular velocities $v_{\rm
  circ}\approx 100 \, \kms$. Note, however, that in DM simulations it has been
  usually assumed that each dark matter halo hosts at least one galaxy. This
  may not be true. Physical processes of galaxy formation are not well known
  and there could be processes that strongly suppress the formation of stars
  inside small haloes, which collapse relatively late in voids. This issue
  will be discussed in detail in the following sections.
  \\

  There is a lot of interest in the void phenomena. However, there is no clear
  definition of a void. Almost every study uses its own definition of a void
  and its own void finder. Recently, the different approaches have been
  compared within the Aspen-Amsterdam Void Finder Comparison Project
  \cite{Joergetal2008}. Here we use a rather simple and direct definition. In
  order to identify voids, we start with a selection of point like objects in
  3D. These objects can be haloes above a certain mass or a certain circular
  velocity or galaxies above a certain luminosity. Thus the voids are
  characterised by the threshold mass or luminosity. The void finding
  algorithm can be applied both to numerical and observational data. In
  numerical simulations it takes into account periodic boundary conditions.
  At first it searches the largest empty sphere which is completely inside the
  given volume. The radius of this empty region is $R_{\rm void}$. Then we
  extend this region to an empty region with irregular shape. We define this
  empty region as union of all spheres with radius $r_{\rm ext} \ge 0.789 \,
  R_{\rm void}$ (thus with a volume greater or equal half of the volume of
  the original void sphere) which can be moved to their place without crossing
  any object or the box boundary. To find the other voids we repeat this
  procedure however taking into account the previously found voids. We now
  search the largest empty sphere which is completely inside the volume and
  does not intersect with any already known void. The voids are characterised
  by volume, shape and orientation.
  \\

  For the re-simulation we selected a void region from a larger periodic
  computational box. To construct suitable initial conditions, at first we
  replaced in a box of side-length $50 \hMpc$ with $2048^3$ particles the low
  mass particles by $256^3$ massive ones. After running the simulation at this
  low mass resolution we determined a void region of interest and, finally, we
  were rerunning the simulation with the original set of particles inside the
  void ($2048^3$) and replacing only outside the void the low mass particles
  by more massive ones.


\section{The baryon content of dwarf galaxy haloes}

  The number density of dwarf galaxies observed in the Universe is apparently
  smaller than the predicted number density of low mass dark matter haloes
  \cite{bosch:07}. The missing galaxies are either not formed because the
  corresponding dark matter haloes do not exist, or no stars are formed in
  small haloes. As shown by Hoeft et al. \cite{hoeft:06} and Okamoto et al.
  \cite{okamoto:08}, the optically thin UV-background already causes baryon
  deficiency in small haloes. In this section we determine the mass scale below
  which photo-heating causes baryon deficiency.

\subsection{Mass accretion history}  

\begin{figure*}[t]
  \begin{center}
  \includegraphics[width=0.9\textwidth,angle=0]{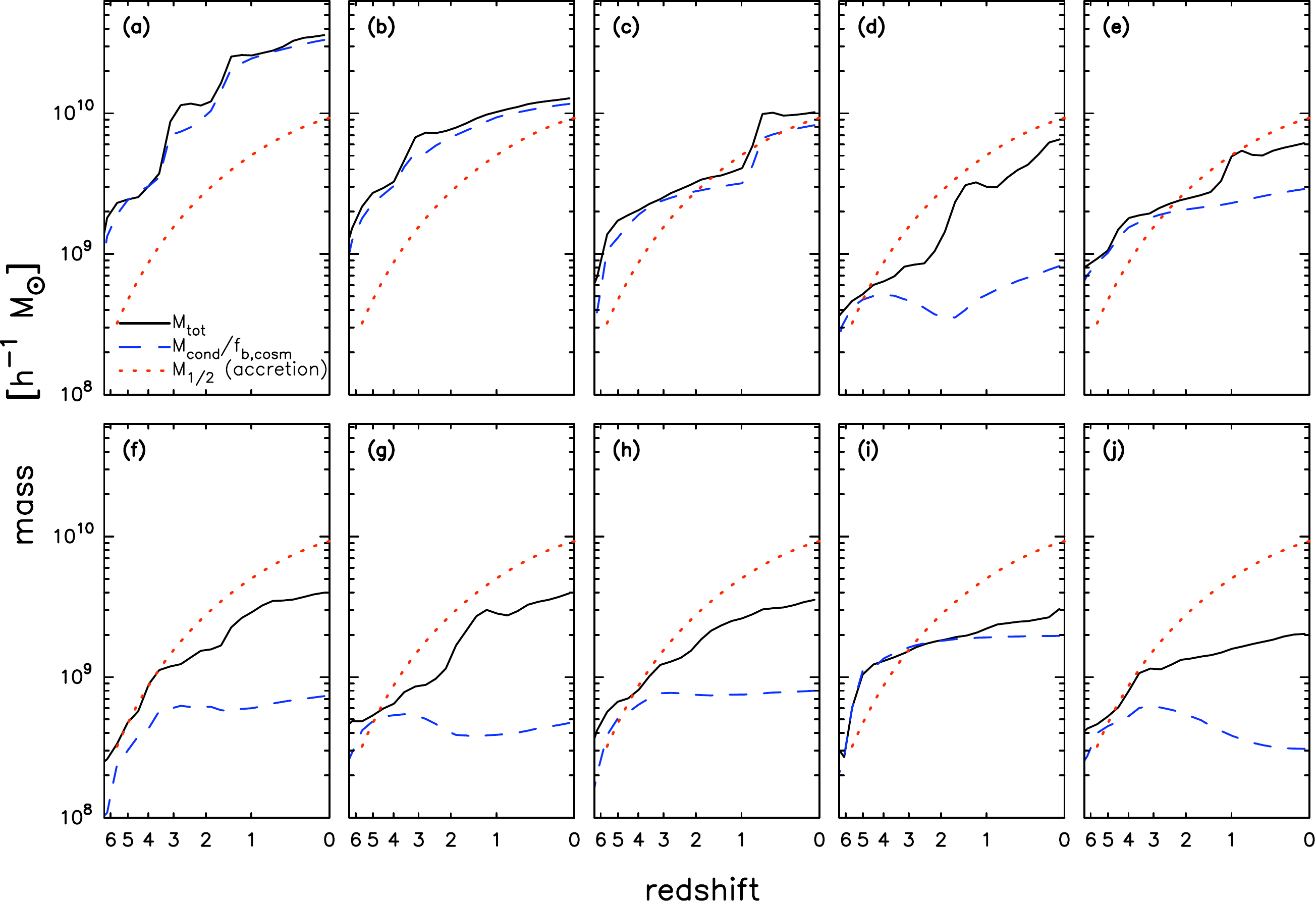}
  \end{center}
  \vspace{-0.3cm}
  \caption
  { 
  The mass accretion histories for several haloes. In each panel the total
  mass (solid line) and the condensed mass (dashed line) of one halo is
  shown. We have divided the condensed mass by the mean cosmic baryon
  fraction, $\langle f_{\rm b}\rangle$, to highlight when the condensed
  mass does not follow the expectation according to the mean baryon
  density. In addition, the characteristic mass (dotted line) is shown as
  derived from baryon fraction in newly accreted matter. One can clearly
  see that haloes cease to condense baryons when their total mass falls
  below the characteristic mass. 
  }
  \label{fig-accretion-histories}
\end{figure*}

  The gravitationally bound structures, \ie galaxies and clusters of galaxies,
  grow in two ways: they accrete surrounding matter and they merge with other
  galaxies or clusters. Numerical simulations allow us to follow the merging
  history of the haloes. Every massive halo at the end of the simulation has a
  large number of progenitors which subsequently merge to the final structure.
  \\

\begin{table}
   \begin{center}
  \begin{tabular}{l|rr}
  refined     & simulation    & mass resolution \\
  region      & bos size      & (dark matter) \\[.7ex]
  \hline
  void           &  $ 50 \hMpc $  & $1.03\times 10^6\:\hMsun $\\
  filament       &  $ 80 \hMpc $  & $8.24\times 10^6\:\hMsun $\\
  \end{tabular}
  \end{center}
  \vspace{-0.3cm}
  \caption{
  Main characteristics of the two simulations analysed for
  determining the characteristic mass.
  }
\end{table}

  We compute the mass accretion history for all haloes in our two simulations,
  see Tab.\,1. To this end we determine the haloes in each simulation snapshot.
  Then we go backward from the final snapshot and search for the most massive
  progenitor in the previous snapshot. Since we have stored more than one
  hundred snapshots for each simulation we can simply search for the most
  massive progenitor in the vicinity of a given halo. However, during merger
  events the halo finder sometimes fails to identify the halo at all, since
  two approaching haloes appear spuriously gravitationally unbound.
  \Fig{fig-accretion-histories} shows the mass accretion histories of ten
  sample haloes, where the jumps are caused by merger events.


\subsection{Characteristic mass}

\begin{figure*}[t]
  \begin{center}
  \includegraphics[width=0.9\textwidth,angle=0]{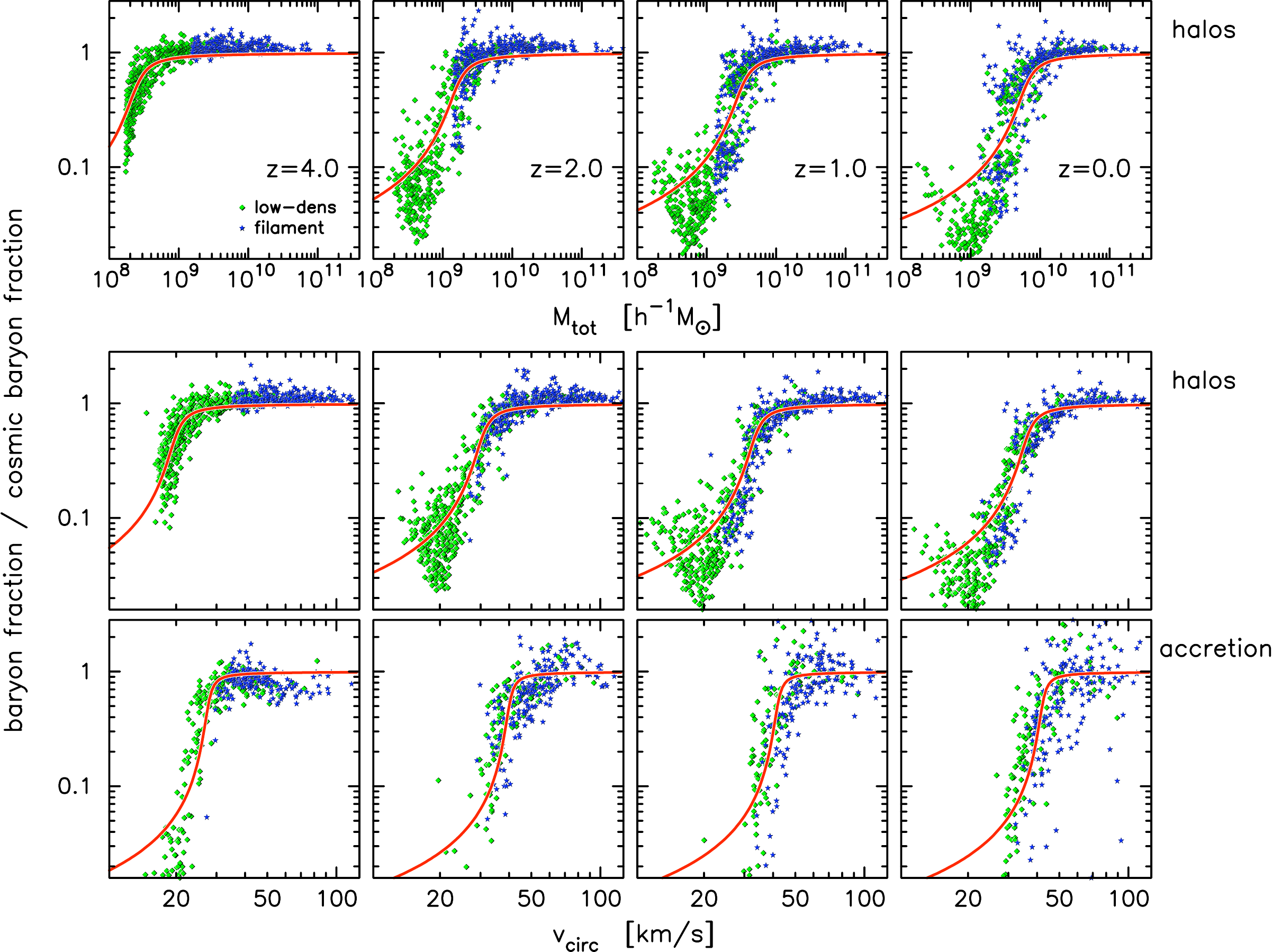}
  \end{center}
  \vspace{-0.3cm}
  \caption
  {
  The baryonic mass fraction in units of the cosmic mean, $\Omega_{\rm
  b}/\Omega_{\rm m}$. Filled squares refer to the simulation of low
  density region while stars refer to the filament region. Only haloes with
  more than 180 dark matter are taken into account. The top row shows
  the baryon fraction at different redshifts as a function of total halo
  mass. The middle row gives the baryon fraction as a function of maximum
  circular velocity. The bottom row shows the baryon fraction of newly
  accreted matter (see text for a more detailed description).
  }
  \label{fig-baryon-fraction}
\end{figure*}

  We determine for all distinct haloes in our simulations the baryon fraction
  within the virial radius, $f_{\rm b} = M_{\rm bar}/M_{\rm tot} = (M_{\rm
  star} + M_{\rm gas})/(M_{\rm star}+M_{\rm gas}+M_{\rm dm})$, where $M_{\rm
  bar}$ denotes the baryonic mass, $M_{\rm tot}$ the total mass, $M_{\rm dm}$
  the dark matter mass, $M_{\rm gas}$ the gas mass and $M_{\rm star}$ in
  stellar mass. We take those haloes into account which consist of more than
  180 dark matter particles. The resulting lowest halo mass is $1.9\times
  10^8\:\hMsun$ and $1.5\times 10^9\:\hMsun$ for the void and for the
  filament region, respectively. Moreover, we exclude all haloes which have
  been within the virial radius of a more massive halo in the past, because
  the dark matter and baryonic masses may be affected by stripping.
  \Fig{fig-baryon-fraction} shows the resulting baryon fraction for both the
  low density region simulation the filament region simulation. The low mass
  haloes evidently have a baryon fraction much smaller than the cosmic baryon
  fraction. Since the transition from `cosmic mean baryon fraction' to `baryon
  deficient' occurs within a rather small mass range this phenomenon can be
  characterised by a single parameter, namely the mass at which the baryon
  fraction amounts in average to half of the cosmic mean, \Mchar.  We
  determine the characteristic mass computing the average baryon fraction in
  mass bins and approximating the resulting curve by the following analytic
  expression
  \begin{equation}
    \begin{split}
    f_{\rm b}
    = &
    \langle f_{\rm b} \rangle \:
    \left\{
      \frac{1}{2}
  	  +
  	  \right. 
  	  \\
  	 & \left.
  	  \frac{1}{\pi}
  	  \arctan
  	  \left(
  	    \frac{\log M_{\rm tot} - \log \Mchar}
  	         {w_M}
      \right)
    \right\}
    .
    \end{split}
    \label{eq-charact-mass}
  \end{equation}  
  We use a different expression than in Ref.\,\cite{hoeft:06} to make the
  fitting procedure more robust, in particular when a second free parameter is
  included, here the width $w_M$. Moreover, we expect that the halo gas, which
  is present independently from the halo size, leads to a floor in the baryon
  fraction. As we have shown in Ref.\,\cite{hoeft:06} the characteristic mass
  depends very little on the mass resolution used in our simulations. This
  indicates that the physical processes causing the baryon deficiency are
  reasonably resolved even if there are only a few tens of gas particles in
  small mass haloes.
  \\

  In Ref.\,\cite{hoeft:06} we focussed on the baryon content of dwarf galaxies
  in cosmological voids. The result we obtained might be affected by
  restricting to void dwarf galaxies. To investigate also the effect of the
  cosmological environment we include here dwarf galaxies in a cosmological
  filament. These galaxies have a higher merger rate than void dwarf galaxies
  with a similar mass but they do accrete less diffuse gas, see
  Ref.\,\cite{fakhouri:10} and references therein. We have excluded all
  galaxies which have passed through the halo of a more massive galaxy since
  their dark matter and baryonic mass could be affected by stripping. Despite
  the cosmologically quite different environments, the characteristic mass of
  both regions coincide well, see \Fig{fig-baryon-fraction} (top row). In
  conclusion, the general mass accretion history of a dwarf galaxy has only
  little effect on their baryon content, \ie there is universal characteristic
  mass for all dwarf galaxies. Note, this result is obtained with a
  homogeneous UV-background. We will argue later that the UV-heating may have
  a crucial impact on the baryon accretion in dwarf galaxies.
  \\

  Observationally, the total mass of a galaxy is difficult to determine.
  Galaxies are preferably characterised by the maximum circular velocity,
  \vmax. One should note that also \vmax is difficult to measure in dwarf
  galaxies since most rotation curves do not approach a maximum, \ie the
  maximum of the circular velocity lies outside the observable gas
  distribution. However, we will characterise haloes mainly by \vmax.
  \Fig{fig-baryon-fraction} (middle row) shows the baryon fractions in
  distinct haloes as a function of the maximum circular velocity.
  \\

  Our aim is to determine what makes haloes below the characteristic mass
  baryon deficient. The baryon content at a given time is accumulated
  over the history of the halo. To assess what prevents at a given
  time haloes from accreting baryonic mass we determine the baryon fraction
  of newly accreted matter
  \begin{equation}
    f_{\rm b}^{\rm accr}
    =
    \frac{ M_{\rm cond}( t_{i+1} ) - M_{\rm cond}( t_{i-1} ) }
         { M_{\rm tot }( t_{i+1} ) - M_{\rm tot }( t_{i-1} ) }
    ,
    \label{eq-def-fb-accr}     
  \end{equation}
  where $t_{i-1}$ and $t_{i+1}$ denote the masses of a given halo at an
  earlier and at a later time, respectively. \Fig{fig-baryon-fraction}
  (bottom row) shows that small haloes do not accrete baryonic mass.
  Since we consider the differential evolution of the dark matter and the
  baryonic masses we have a large scatter in the characteristic mass
  of newly accreted matter. However, similar to the baryon fraction
  in haloes there is a clear separation between low mass haloes which accrete
  virtually no baryonic matter and high mass haloes which accrete matter with
  the mean cosmic baryon fraction.

\subsection{Evolution of the characteristic mass}
  
\begin{figure*}[t]
  \begin{center}
  \includegraphics[width=0.8\textwidth,angle=0]{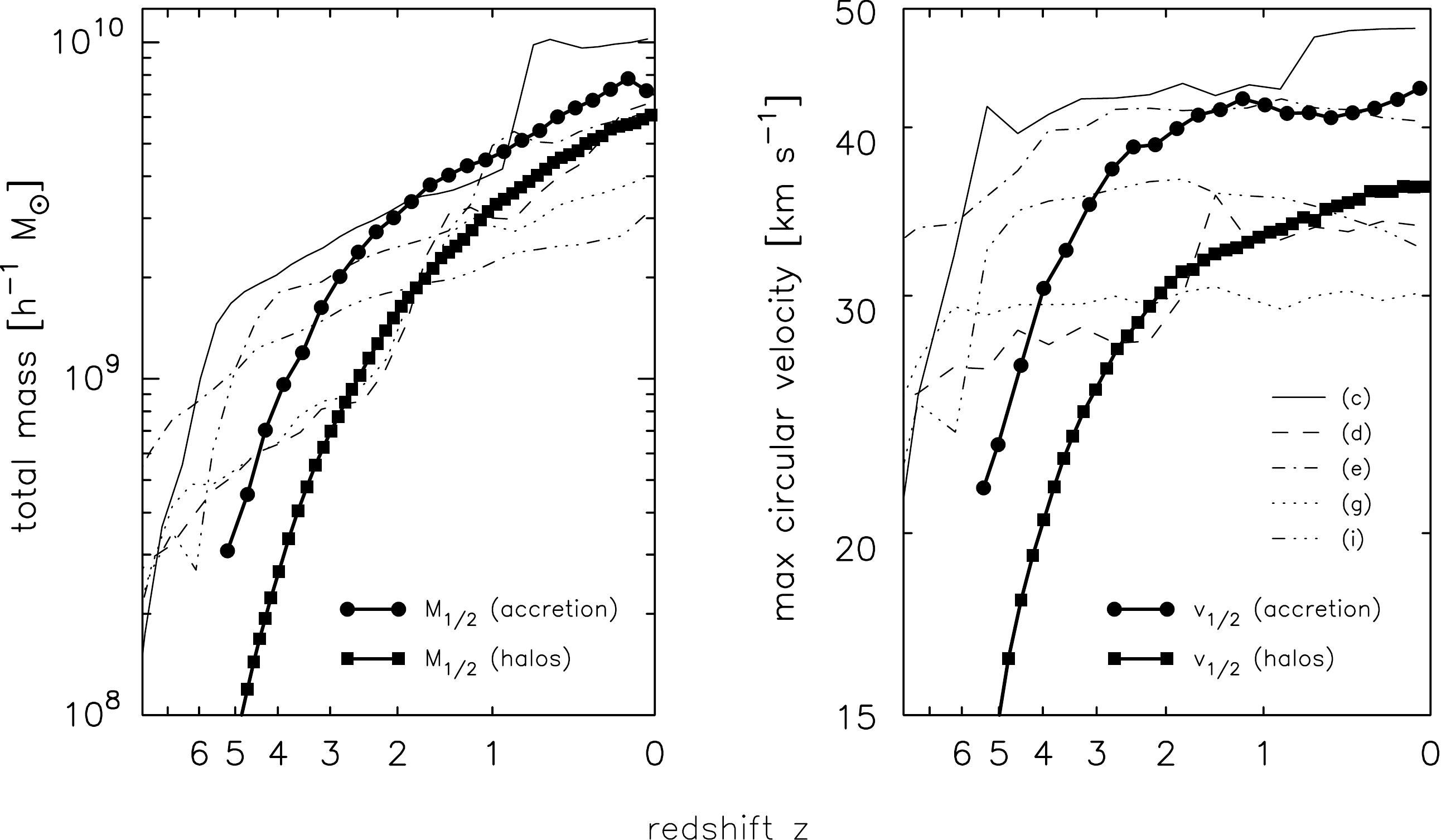}
  \end{center}
  \vspace{-0.3cm}
  \caption{
  {\it Left panel:} Evolution of the characteristic mass of the baryon
  fraction in distinct haloes (squares) and in newly accreted matter
  (circles). For comparison the mass accretion histories of five haloes, as
  depicted in \Fig{fig-accretion-histories}, are shown. {\it Right panel:}
  Evolution of the characteristic circular velocity in distinct haloes and
  in newly accreted matter. Additionally, the circular velocity histories
  of the haloes in the left panel is shown. 
  }
  \label{fig-baryon-fraction-evolution}
\end{figure*}

  \Fig{fig-baryon-fraction} shows that the characteristic mass decreases
  with redshift. For instance, at redshift $z=4$ the characteristic mass
  is about $3\times 10^8\:\hMsun$, while it amounts to $6\times
  10^9\:\hMsun$ at $z=0$. \Fig{fig-baryon-fraction-evolution} shows the
  redshift evolution of the characteristic mass as obtained from combined
  halo lists from our two simulations. For comparison also a few mass
  accretion histories from \Fig{fig-accretion-histories} are shown. 
  \\

  Plotted as a function of circular velocity the evolution of the
  characteristic mass of the newly accreted matter has a rather simple
  behaviour: after a steep increase it is almost constant at $\sim 40 \,\kms$
  for $z\lesssim3$. Note, in Ref. \cite{hoeft:06} we found a lower value since
  we used the circular velocity at the virial radius, $v(r_{\rm vir})$,
  instead of $\vmax$. Moreover, we considered only the baryon fraction in
  haloes. Since semianalytic models of galaxy formation assume a low-mass
  cut-off for gas accretion, see \eg Ref.\,\cite{somerville:02} and
  Ref.\,\cite{li:10}, we investigate here also the characteristic mass of
  newly accreted matter. 
  \\

  Because the circular velocity of most of the dwarf galaxy haloes is almost
  constant with time, except a major merger occurs, one can separate two types
  of evolutionary scenarios for dwarf galaxies: (i) small galaxies with $\vmax
  \ll 40\,\kms$ accrete baryons only for $z \gtrsim 4$ and (ii) galaxies with
  $\vmax \gg 40\,\kms$ are not affected at all. In
  \Fig{fig-accretion-histories} we have also depicted for each halo the
  evolution of $M_{\rm 1/2}^{\rm accr}$ as shown in
  \Fig{fig-baryon-fraction-evolution} (left panel). One can see nicely that
  haloes get baryon deficient when their halo mass is below $M_{\rm 1/2}^{\rm
  accr}$.

\subsection{Can the characteristic mass explain the void phenomenon?}

  The void phenomenon comprises two observational findings \cite{Peebles2001}:
  First, the distribution of massive galaxies ($\geq L_\ast$) shows large
  empty regions, the `voids', see Sec.\, 2.4. Low-luminosity galaxies populate
  the voids but with a low volume density compared to the dense regions
  \cite{Hoyleetal2005}. Secondly, the galaxies found in voids are not special
  at all. According to the concordance $\Lambda$CDM cosmology voids are filled
  by dark matter haloes with the size of dwarf galaxies \cite{gottloeber:03}.
  To judge if this is at odds with the void phenomenon in observations one has
  to assign a luminosity to the dark matter haloes. Recently, Tinker and
  Conroy \cite{tinker:09} applied a Halo Occupations Distribution (HOD) to a
  large dark matter simulation and analysed void regions. They found that
  the haloes in void regions which host galaxies with an absolute magnitude $M_r
  > -10 $ are located close to the walls of the voids. Only even fainter
  galaxies, \ie galaxies which are significantly below the current detection
  limit of surveys covering large void regions \cite{Hoyleetal2005}, populate
  the entire void volume. According to Tinker and Conroy galaxies with $M_r =
  -10 $ reside in haloes with a mass $\sim 10^{10} \hMsun$. Hence, the
  photo-heating discussed here would reduce the luminosity of the haloes
  filling the entire void volume rendering those galaxies even more difficult to
  detect.
  \\

  The HOD used by Tinker and Conroy implies that the mass-to-light ratio
  increases by a factor of ten from haloes with a mass of $10^{13} \hMsun$ to
  haloes with $10^{10} \hMsun$, independent of the environment. This is
  consistent with the compilation of the baryon content of galaxies by McGaugh
  \cite{mcgaugh:10}. Haloes with the mass of $10^{10} \hMsun$ seem to host
  only 5\,\% of the baryonic mass corresponding to the cosmic mean. In our
  simulations we do not find any indication that the UV-background reduces the
  baryon fraction in galaxies significantly more massive than the
  characteristic mass. Therefore, we speculate that stellar feedback plays a
  key role in reducing the baryon fraction galaxies with $M_{\rm tot} > M_{1/2}$. 
  Note that the baryon fractions in Ref.\,\cite{mcgaugh:10} shows indeed
  a break at $v_{\rm circ} \approx 30 \, {\rm km \, s^{-1} } $, consistent
  with our findings for the characteristic mass.
  \\

  In summary, the surprising low number of low-luminosity galaxies in voids
  can be explained by a mass-dependent mass-to-light ratio, as obtained from a
  global Halo Occupation Distribution and from a proper mass modelling of
  galaxies. Haloes above the characteristic mass are expected to reside close
  to the walls of voids. The photo-heating discussed here reduces the
  luminosity of the low-mass and low-luminosity galaxies which can populate
  the entire void volume. Current surveys are limited to detect galaxies close
  to the walls due to the sensitivity limits of the surveys.


\section{Why small haloes fail to accrete gas}

  In the previous section we have shown that dark matter haloes with $\vmax
  \ll 40\,\kms$ do accrete dark matter for $z \lesssim 4$ while they do
  not accrete baryons for smaller redshifts. Halos with $\vmax < 35\,\kms$
  are generally baryon deficient for $z=0$. These results are obtained for
  high-resolution hydrodynamical cosmological simulations including a homogeneous
  photo-heating of an optically thin medium and including radiative cooling
  for a primordial chemical composition. In this section we investigate in
  detail the origin of the baryon deficiency which is evidently related to
  the photo-heating by the UV background. To this end we set up an
  analytic model and investigate how baryons cool in dwarf galaxies.

\subsection{Outline of the analytic model}

  Our simple model is constructed as follows: In the beginning the 
  gas in the universe is homogeneously distributed. In the process of
  structure formation gas is accreted by galaxies, \ie its density increases.

  Starting from a homogenous distribution the gas in the periphery of haloes
  and in the IGM follows the evolution of the dark matter. For our model here
  we assume that there is universal evolution of the overdensity, $\Delta(t,
  \Delta_0)$, which only depends on the time and on the final overdensity, see
  \Fig{fig-delta-evolve}. By the help of \Eq{eq-temp-evolve} we can determine
  the temperature as a function of the overdensity. For a given
  temperature-density relation we can determine the gas density profile of
  NFW-type haloes. Assuming that the gas density in the periphery of the halo
  is given by the mean cosmic baryon fraction we can determine the central gas
  density and temperature. If the resulting cooling time is short the halo
  would accrete baryonic matter, if the cooling time is long the halo would
  fail to accrete baryonic matter.
\\

\subsection{Effective equation of state of the IGM}
\label{sec-eos}

\begin{figure}[t]
  \begin{flushright}
  \includegraphics[width=0.45\textwidth,angle=0]{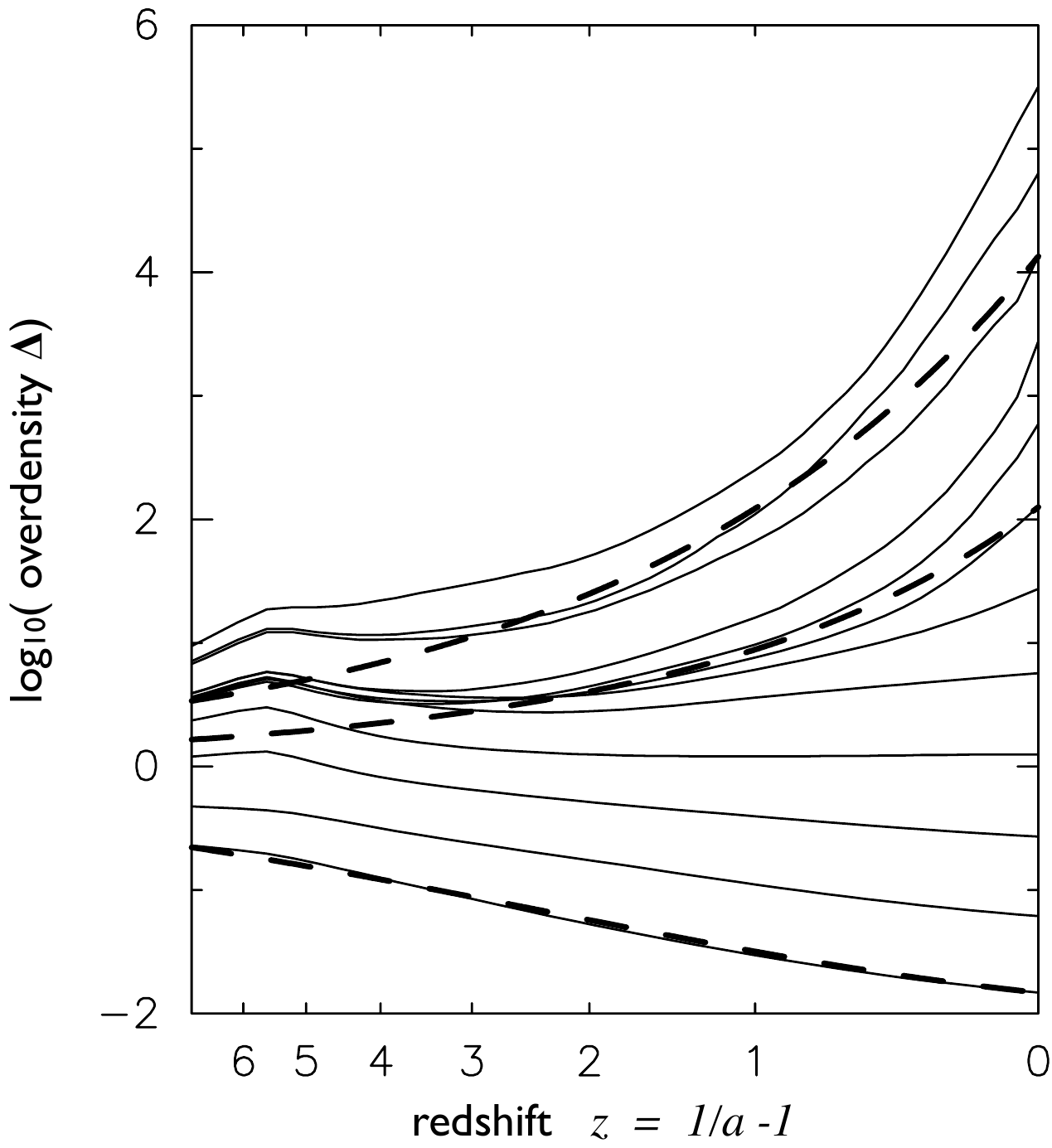}
  \end{flushright}
  \vspace{-0.6cm}
  \caption
  {
  The overdensity evolution $\Delta(a,\Delta_0)$. We group in
  the simulation snapshot for $z=0$ all particles according to their
  overdensity. Then trace back the density of the SPH-particles and
  compute the average density of the groups for earlier redshifts (solid
  lines). Moreover, we show three approximated overdensity evolutions (dashed lines)
  according to \Eq{eq-delta-evolve}. See \Sec{sec-eos} for details. The
  peak-like structures at $z \approx 6$ are caused by photo-evaporation
  and are neglected in our approximation.
  }
  \label{fig-delta-evolve}
\end{figure}

  For a known overdensity evolution, $\Delta(t,\Delta_0)$, we can integrate
  the thermal evolution of the IGM, \Eq{eq-temp-evolve}. To determine the
  heating and the cooling, ${\cal H}(t)$ and $\Gamma(t)$, respectively, we
  have to know the abundance of each species, \ie the ionisation state of
  hydrogen and helium. As in the cosmological simulations we adopt collisional
  equilibrium, hence, we can determine for a given time and a given
  temperature the relative abundances. We solve the rate equations for
  collisional ionisation, for recombination, and for photo-ionisation. Now we
  can compute the photo-heating and cooling of the gas. Time-dependent
  specific energy injection rates, $\epsilon_i$, are tabulated according to
  the UV-background model. As a result the total photo-heating can be written
  as  ${\cal H} = n_{\rm HI}\epsilon_{\rm HI} + n_{\rm HeI}\epsilon_{\rm HeI}
  + n_{\rm HeII}\epsilon_{\rm HeII}$. Cooling rates are used as given in
  \cite{katz:96}. Hence, for a given overdensity evolution,
  $\Delta(t,\Delta_0)$ and UV-background model we can compute the final gas
  temperature, $T(t_0, \Delta_0)$.
  \\

  We use our numerical simulations to estimate the overdensity evolution,
  $\Delta(t,\Delta_0)$. In Gadget the hydrodynamical equations are solved
  by the SPH method, hence, the fluid is represented by mass elements. The
  SPH-particles can be traced through the simulation. We group
  SPH-particles according to their overdensity at the end of the
  simulation, $\Delta_0$, regardless if they are located in a galaxy, in a
  cosmological filament or in the field. Then, we trace back the density
  of the particles and compute the average density of each group, see
  \Fig{fig-delta-evolve}. For the integration of the thermal evolution of
  the IGM we approximate the overdensity evolution by an analytic
  expression, namely
  \begin{equation}
    \log \Delta(t,\Delta_0) 
    = 
    \log  \Delta_0  
    \cdot
    \left|
      \frac{G(a,\Delta_0)}
           {G(1,\Delta_0)}
    \right|
    ,
    \label{eq-delta-evolve}
  \end{equation}
  where we use the cosmic expansion factor, $a = 1 / (z+1)$, as time
  variable. We approximate the time evolution by $G(a,\Delta_0) =
  |(a+1)^{m(\log\Delta_0)}-1 |$, and determine for each final overdensity
  the exponent $m(\log\Delta_0)$. We fit \Eq{eq-delta-evolve} to each
  curve in \Fig{fig-delta-evolve}, where we take only the data for $z \leq
  3$ into account. In the overdensity range $-2 < \log\Delta_0 < 6 $ the
  resulting exponents are reasonably reproduced by $m(\log\Delta_0)=
  1.3\log\Delta_0- 0.26\,(\log\Delta_0)^2$. This finally provides an
  analytic model for the overdensity evolution, $\Delta(t,\Delta_0)$.
  \\

\begin{figure}[t]
   \begin{flushright}
   \includegraphics[width=0.45\textwidth,angle=0]{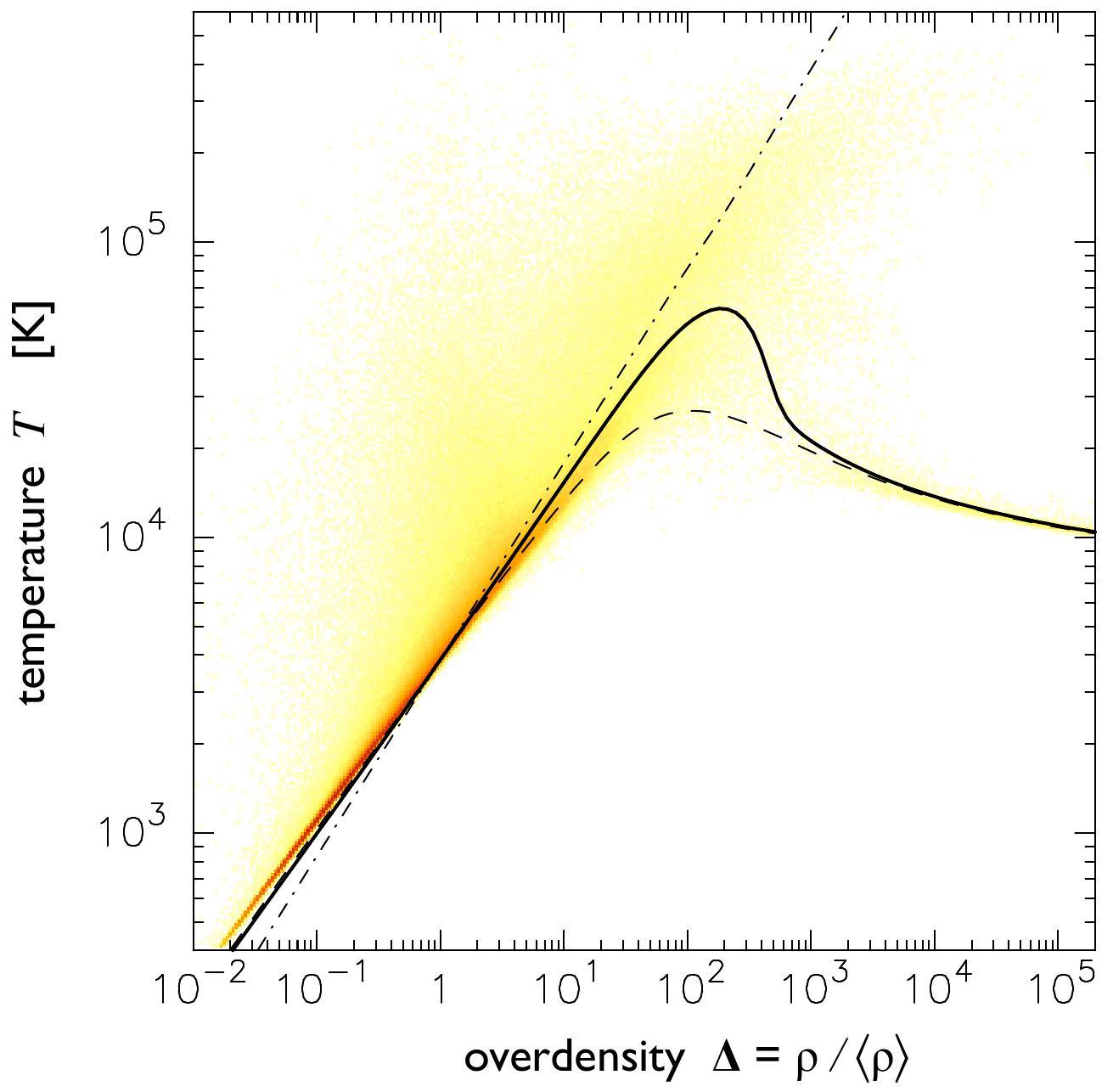}
   \end{flushright}
   \vspace{-0.6cm}
   \caption
   {
   Temperature-density relation, $T(\Delta,t_0)$ for the overdensity
   histories as given in \Eq{eq-delta-evolve} (thick solid line). In
   addition two extreme scenarios for the overdensity history are shown.
   First, all gas is compressed or decompressed to its final overdensity,
   $\Delta_0$, at the very beginning, \ie before reionisation takes place
   (dashed-dotted line). Secondly, all gas evolves at cosmic mean density
   until $z=0$ and is then instantaneously compressed or decompressed to
   its final overdensity. The latter clearly results in an adiabatic
   relation (dash-dotted line). Finally, the distribution of the
   SPH-particles in the simulation is shown. 
   }
   \label{fig-Tn-delta-evolve}
 \end{figure}

  In \Fig{fig-Tn-delta-evolve} the temperature-density relation,
  $T(\Delta,t_0)$, obtained by integrating \Eq{eq-temp-evolve} using the
  analytic formula for the overdensity evolution described above, is
  shown. For comparison the density-temperature phase-space, as obtained
  from the numerical simulation, is also depicted. For high densities,
  $\Delta \gtrsim 10^3$, the analytic result, $T(\Delta,t_0)$, and the
  distribution of the SPH-particle coincides very well, because the
  thermal state is determined by thermal equilibrium. For low densities,
  most of the SPH-particles also follow $T(\Delta,t_0)$. Finally, a
  significant fraction of the SPH-particles has temperatures substantially
  above $T(\Delta,t_0)$, since in the numerical simulation particles are
  also heated by dissipation at shock fronts.
  \\

  Which impact has the overdensity evolution? To answer this question we
  investigate two extreme scenarios. First, gas stays at mean cosmic
  density until $z=0$ and is then instantaneously compressed or
  decompressed to the final overdensity. This clearly leads to an
  adiabatic temperature-density relation, see \Fig{fig-Tn-delta-evolve}
  (dash-dotted line). Secondly, gas is brought to the final overdensity
  before reionisation takes place and has then constant overdensity. The
  resulting temperature-density (dashed line) relation is only slightly
  different from the result for the average cosmic density evolution
  (solid line); the transition to thermal balance occurs at lower
  overdensities, since the compressed gas has more time to cool.
  \\

  For the overdensity history introduced in \Eq{eq-delta-evolve} we find
  for low overdensities, $\Delta \lesssim 100$, an effective
  equation-of-state, \ie the temperature-density relation follows a
  power-law,
  \begin{equation}
    T_{\rm eos} (\Delta,t=t_0)
    = 
    T_0\:
    \Delta^\alpha  
    ,
    \label{eq-eff-eos}
  \end{equation}  
  with $T_0 = 3.85 \times 10^3 \:{\rm K}$ and $\alpha = 0.59$. 
  \\

\begin{figure}[t]
  \begin{flushright}
  \includegraphics[width=0.45\textwidth,angle=0]{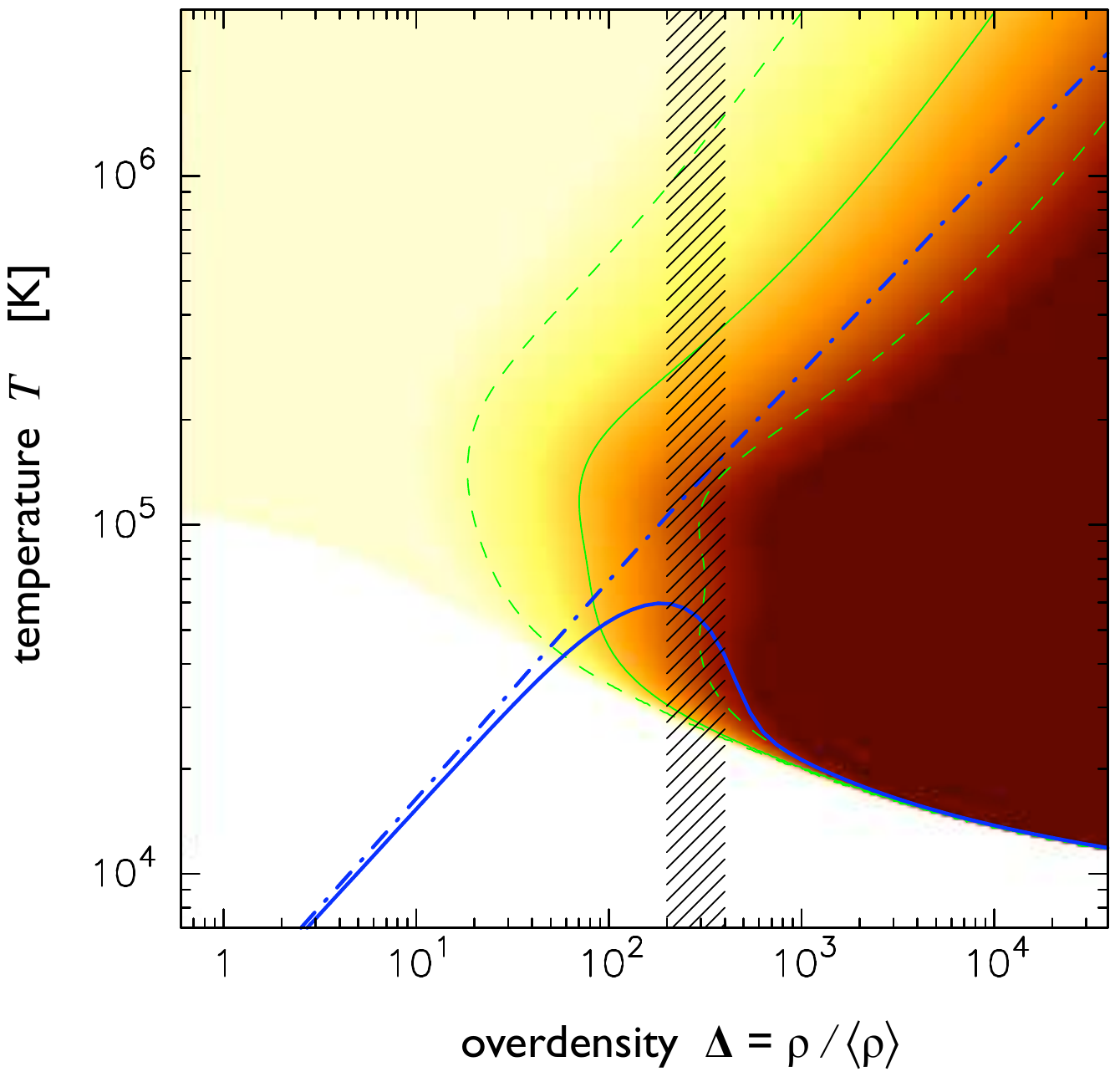}
  \end{flushright}
  \vspace{-0.6cm}
  \caption
  {
  Theoretical result for $T(\Delta)$ with cooling (solid line) and without
  (dash-dotted line). The result without cooling follows the power-law $T
  = 3.85 \, {\rm K} \: (\delta+1)^{0.595}$. Cooling times are indicated by
  colours and contour lines. Contours from left to right are at 10x Hubble
  time (dashed line), Hubble time (solid line), and 0.1x Hubble time
  (dashed line). Cooling times are also indicated by colours, the blank
  area indicate the regime of UV heating. The hashed area highlights the
  transition region from cumulative photo-heating to thermal balance. 
  }
  \label{fig-Tn-theo}
\end{figure}

  By plotting the cooling times in the temperature-density phase-space we
  can clearly see at which overdensities the gas is in thermal balance,
  see \Fig{fig-Tn-theo}. For low overdensities, $\Delta\lesssim 10^2$,
  cooling has only a negligible effect, \ie the thermal state of the gas
  is determined by the cumulative photo-heating. In contrast, at high
  overdensities, $\Delta\gtrsim 10^3$, the temperature lies exactly at the
  boundary between heating and cooling, \ie it is in thermal balance.
  Therefore, in the range $10^2 < \Delta < 10^3$ the transition from the
  pure photo-heating to the equilibrium between heating and cooling takes
  place.

\subsection{The halo density profile}
\label{sec-central-gas-density}

  We now set up the second part of our toy model, namely, the spherical
  halo model, with the aim to estimate the innermost gas density as a
  function of halo mass.
  \\
 
  For a spherically symmetric halo in hydrostatic equilibrium  
  the pressure force and the gravitational force are in balance,
  \begin{equation}
    \frac{1}{\rho}
    \frac{{\rm d} P}{ {\rm d} r }
    =
    -
    \frac{G M_{\rm tot} (<r)}{ r^2 }
    \label{eq-hydro-equi}
    ,
  \end{equation}
  where $G$ is the gravitational constant and $M_{\rm tot}(<r)$ is the
  total mass enclosed by the radius $r$. Assuming an effective
  equation-of-state, as given in \Eq{eq-eff-eos}, we can write the
  left-hand side of \Eq{eq-hydro-equi} in terms of the gas overdensity. To
  this end we expand the gas density, $\rho = \Delta \, n_{\rm b}
  m_{\rm p}$, where $m_{\rm p}$ is the proton mass. The pressure
  can be written as $P = \Sigma_i n_i k_{\rm B} T = (\Delta n_{\rm
  b}  / \mu ) \, k_{\rm B} T_0 \Delta^\alpha$, where we have used
  the molecular weight, $\mu = n_{\rm b} / \Sigma_i n_i$, and
  \Eq{eq-eff-eos}. Now we rewrite the left-hand side of \Eq{eq-hydro-equi},
  \begin{equation}
    \frac{k_{\rm B} T_0}
         {\mu m_{\rm p} }
    (\alpha+1) \:
    \Delta^{\alpha-1} \:     
    \frac{{\rm d}\Delta}
         {{\rm d}r}
    .
    \label{eq-lhs-hydro-equi}
  \end{equation}
  \\

  We simplify the right-hand side of \Eq{eq-hydro-equi} by assuming that
  the distribution of the total mass is determined by the dark matter. For
  the density profile of the latter we follow Ref.\,\cite{navarro:97b},
  \begin{equation}
    \rho_{\rm dm}
    =
    \rho_{\rm s}
    \frac{4}{x \, (1+x)^2 }
    ,
    \label{eq-nfw-profile}
  \end{equation}
  with $x = r/r_{\rm s}$. The scaling radius, $r_{\rm s}$, is related to
  the virial radius, $r_{\rm vir}$, of a halo by $r_{\rm s} = r_{\rm vir}
  / c $, where $c$ is the concentration parameter. The mass enclosed by
  the scaled radius $x$ is
  \begin{equation}
    \begin{split}
    M(<x)
    & = 
    16 \pi \, r_{\rm s}^3 \rho_{\rm s} \:
    \left\{
    	\ln(1+x) - \frac{x}{1+x}
    \right\}
    \\
    & = 
    16 \pi \, r_{\rm s}^3 \rho_{\rm s} \: F(x)
    .
    \end{split}
    \label{eq-nfw-cum-mass}	
  \end{equation}
  Evidently, the virial radius encloses the virial mass, $M_{\rm vir} =
  M(<c)$. Dividing \Eq{eq-nfw-cum-mass} by the virial mass leads to
  \begin{equation}
    M(<x)
    =
    M_{\rm vir} \:
    F(x) / F(c)
    .
  \end{equation}
  Using $r = r_{\rm vir} x / c$ and the expression \Eq{eq-lhs-hydro-equi}
  we can rewrite \Eq{eq-hydro-equi}
  \begin{equation}
    \frac{k_{\rm B} T_0 }
         {\mu m_{\rm p} }
    \:
    (\alpha + 1 ) \:
    \Delta^{\alpha - 1}
    \:
    {\rm d} \Delta
    =
    \frac{ G M_{\rm vir} }
         { r_{\rm vir} }
    \frac{ F(x)  }
         { F(c)  }
    \frac{ c }
         {x^2}
    \:
    {\rm d}x
    .
    \label{eq-balance-hydro-grav}
  \end{equation}
  We can further simplify the expression on the right-hand side by
  introducing the maximum circular velocity. The circular velocity is
  given by
  \begin{equation}
    \begin{split}
    v_{\rm circ} (r)
    & = 
    \sqrt{
    	\frac{G M(<r)}
    	     {r}
    }
    \\
    & = 
    \sqrt{
    	\frac{ G M_{\rm vir} }
    	     { r_{\rm vir} }
    	\frac{c}
    	     {x}
    	\frac{F(x)}
    	     {F(c)}
    }     
    .
    \end{split}
  \end{equation}
  The quotient $F(x)/x$ is independent from the halo properties and has a
  maximum value of $\sim 0.22$. Therefore, we can write for the maximum
  circular velocity
	 \begin{equation}
		v_{\rm max}^2
		=
		\frac{ G M_{\rm vir} }
			 { r_{\rm vir} }
		\frac{0.22 \,c}
			 {F(c)}
		.    
		\label{eq-vmax-Mvir-rvir}
	  \end{equation}
  \Eq{eq-balance-hydro-grav} is now independent from the concentration
  parameter of the halo, namely
	\begin{equation}
	  \frac{k_{\rm B} T_0 }
		   {\mu m_{\rm p} }
	  \:
	  (\alpha + 1 ) \:
	  \Delta^{\alpha - 1}
	  \:
	  {\rm d} \Delta
	  =
	  \frac{v_{\rm max}^2}
		   {0.22} \:
	  \frac{ F(x) }
		   {x^2}
	  \:
	  {\rm d}x
	  .
	  \label{eq-balance-hydro-grav-2}
	\end{equation}
  To determine the innermost gas overdensity we integrate
  \Eq{eq-balance-hydro-grav-2} starting from the periphery of the halo,
  $x=x_{\rm out}$, to the center, $x=0$. Since the halo gas is highly
  ionised we adopt $\mu=0.59$. The resulting central overdensity is 
	\begin{equation}
	  \begin{split}
	  \Delta_{\rm centre}^\alpha
	  =
	  \left( 
		\frac{ T_0 }
			 { 10^4 \: {\rm K } } 
	  \right)^{-1} \:
	  \left( 
		\frac{ v_{\rm max} }
			 { 5.6 \: {\rm km \, s^{-1} } } 
	  \right)^{2} \:   
	  \\
	  \times \frac{ \alpha }
		   { (\alpha + 1 ) } \:
	  \left( 
		1 - \frac{ \ln(1+x_{\rm out}) }
				 {x_{\rm out}}
	  \right) 
	  \;\; + \;\;
	  \Delta_{\rm out}^{\alpha} 
	  ,
	  \end{split}
	  \label{eq-central-gas-density}
	\end{equation}
  where $\Delta_{\rm out}$ is the gas overdensity at $x_{\rm out}$. We
  estimate the outer overdensity by assuming that it is identical to the dark
  matter overdensity.
  \\

  From \Eq{eq-nfw-profile} we get
	\begin{equation}
	  \Delta_{\rm dm}
	  =
	  \frac{ \rho_{\rm dm} }
		   { \langle \rho_{\rm dm} \rangle }
	  =
	  \frac{ \rho_{\rm s} }       
		   { \langle \rho_{\rm dm} \rangle }
	  \frac{ 4 }
		   { x ( 1+x)^2 }
	  .     
	  \label{eq-dm-overdensity}
	\end{equation}
  The virial radius is determined by the cumulative overdensity, $\Delta_{\rm
  c}(z)$, at which virialisation is expected. We use the approximation of
  Ref.\,\cite{bryan:98} for $\Delta_{\rm c}(z)$. For the virial mass we write
  now 
	\begin{equation}
	  M_{\rm vir}
	  =
	  \frac{4}{3}
	  \pi
	  r_{\rm vir}^3
	  \Delta_{\rm c}
	  \langle \rho_{\rm dm} \rangle
	  =
	  16 \pi 
	  r_{\rm s}^3
	  \rho_{\rm s}
	  F(c)
	  ,
	  \label{eq-Mvir-Deltac}
	\end{equation}
  where we have used \Eq{eq-nfw-cum-mass} for the last equality.
  Substituting $r_{\rm s}/\langle \rho_{\rm dm} \rangle$ we can write for
  \Eq{eq-dm-overdensity}
	\begin{equation}
	  \Delta_{\rm dm} (x_{\rm out})
	  =
	  \frac{\Delta_{\rm c}(z)}
		   {3} \:
	  \frac{c^3}
		   {F(c)} \:
	  \frac{1}
		   {x_{\rm out} ( 1 + x_{\rm out} )^2 }
	  ,
	  \label{eq-outer-overdensity}
	\end{equation}
  which allows us to estimate $\Delta_{\rm out}$ in \Eq{eq-central-gas-density}.

  We wish to give the central gas overdensity as a function of the maximum
  circular velocity. To utilise \Eq{eq-outer-overdensity} we have to find a
  relation between $v_{\rm max}$ and the concentration parameter. Maccio {\it
  et al.} \cite{maccio:07} give an relation between $c$ and the virial mass of
  a halo for $z=0$,
	\begin{equation}
	  c(M_{\rm vir})
	  =
	  10.47 \times
	  \left(
		 \frac{ M_{\rm vir} }
			  { 10^{12} \: h^{-1} M_\odot }
	  \right)^{-0.109}
	  .
	  \label{eq-c-M-relation}
	\end{equation}
  From our simulations we find that $M_{\rm vir}$ and the maximum circular
  velocity are on average related by
	\begin{equation}
	  \frac{ M_{\rm vir} }
		   { 10^{12} \: h^{-1} M_\odot }
	  =
	  \left(
		\frac{ v_{\rm max} }
			 { 210 \: {\rm km \, s^{-1} } }
	  \right)^{2.9}
	  .
	\end{equation}
  Moreover, the concentration parameter is expected to scale with redshift
  according to $c \sim (1+z)^{-1}$ \cite{zhao:03,wechsler:02}. Thus, we
  obtain an relation between concentration parameter and $v_{\rm max}$,
  namely
	\begin{equation}
	  \begin{split}
	  & c(v_{\rm max})
	  = \\
	  & \quad 10.47 \:
	  \left(
		\frac{ v_{\rm max} }
			 { 210 \: {\rm km \, s^{-1} } }
	  \right)^{-0.32} \:
	  \frac{1}
		   {(z+1)}
	  ,
	  \end{split}
	  \label{eq-rel-c-vmax}
	\end{equation}
  which allows us finally to evaluate \Eq{eq-central-gas-density}.
  \\

\begin{figure}[t]
  \begin{flushleft}
  \includegraphics[width=0.45\textwidth,angle=0]{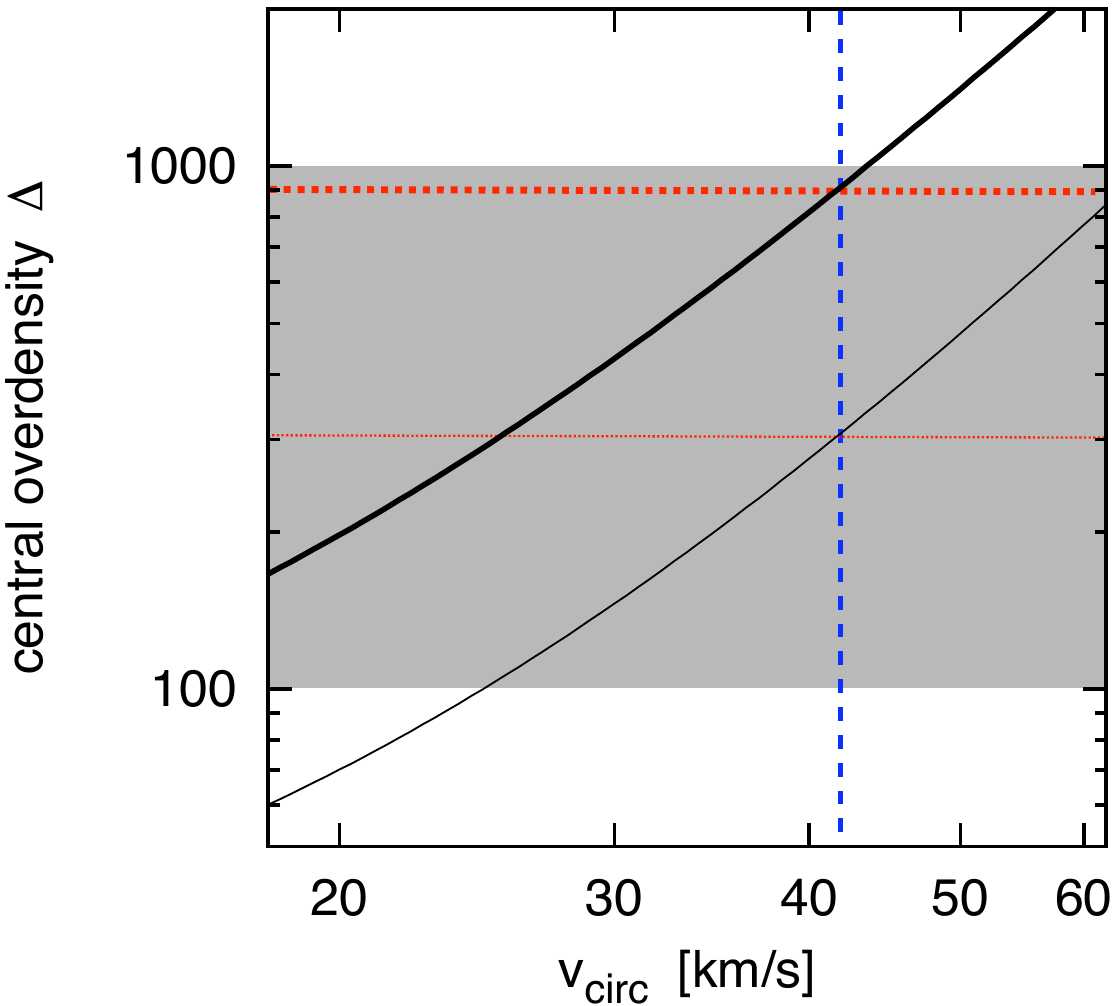}
  \end{flushleft}
  \vspace{-0.6cm}
  \caption
  {
  Central gas overdensity as a function of the halo circular velocity. We
  compute the central overdensity according to \Eq{eq-central-gas-density}
  (thick solid line). Moreover, we plot the overdensity at the scaling radius,
  $r_{\rm s}$ (thin solid line). The grey area indicates the transition region
  from pure UV heating to thermal balance, see \Sec{sec-eos} or
  \Fig{fig-Tn-theo}. The blue dashed line indicates the characteristic
  circular velocity obtained from the numerical simulations. The thick and
  the thin red dotted lines indicate a condensation criterion of $\Delta =
  900$ and 300, respectively.
  }
  \label{fig-central-overdensity}
\end{figure}

  Finally, we compute the central gas overdensity as follows: For a given
  circular velocity we determine the concentration parameter of the halo
  by \Eq{eq-rel-c-vmax}. Then we estimate the outer gas density, here we
  assume that $\Delta(x_{\rm out}) = \Delta_{\rm dm}(x_{\rm out})$, see
  \Eq{eq-outer-overdensity}, at three times the virial radius, $x_{\rm
  out} = 3c$. Finally, we compute the central gas overdensity,
  \Eq{eq-central-gas-density}, as a function of the circular velocity, see
  \Fig{fig-central-overdensity}. 
  \\

  To estimate the characteristic circular velocity by our analytic model
  we have to give a criterion at which central overdensity gas condenses
  in a halo centre. The transition from the effective equation-of-state to
  the thermal balance takes place in the overdensity range of $10^2 <
  \Delta < 10^3$. For an overdensity of 900 our analytic model would match
  the characteristic circular velocity obtained from the numerical
  simulations. However, the resolution in the numerical simulation is not
  sufficient to reach the innermost gas density as assumed in our model.
  To mimic the numerical resolution we integrate
  \Eq{eq-central-gas-density} only up to the scaling radius, see thin line
  in \Fig{fig-central-overdensity}. In this case an overdensity of 300
  corresponds to the characteristic circular velocity. This leads to two
  conclusions: First, the analytic model matches the numerical result very
  well. Secondly, better resolved simulations may result in a lower
  characteristic velocity. However, gas mixing in the halo centre may
  generally prohibit an ideal hydrostatic equilibrium.

\subsection{Evolution of the characteristic mass}

\label{sec-Mchar-evolv}

\begin{figure*}[t]
  \begin{center}
  \includegraphics[width=0.8\textwidth,angle=0]{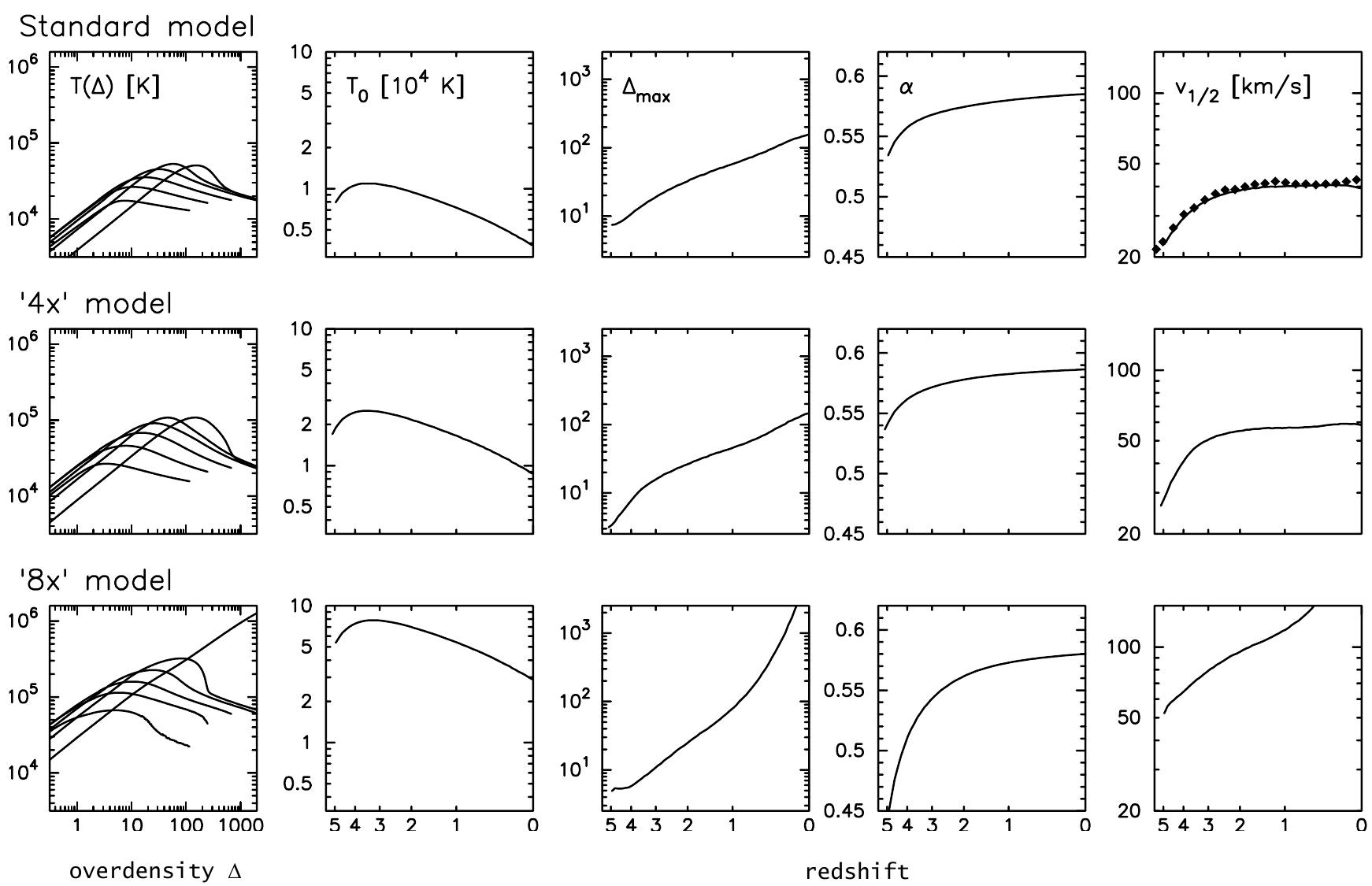}
  \end{center}
  \vspace{-0.6cm}
  \caption
  {
  {\it Left panels}: The density-temperature $T(\Delta,t)$ for several redshifts and
  three different UV-background models. The next three columns show the 
  redshift evolution of the parameters of $T(\Delta,t)$: The normalisation
  and the effective e.o.s., $T_0$, the position of the maximum, $\Delta_{\rm max}$, 
  and the slope of the e.o.s., $\alpha$. Finally, for each UV-background model
  the resulting characteristic circular velocity is shown, 
  see \Eq{eq-central-gas-density-simple}. The upper right panel shows also
  the characteristic velocity obtained from the numerical simulations.
  }
  \label{fig-Tn-redshift-evolution}
\end{figure*}

  In the previous section we have shown that our model
  well reproduces the characteristic circular velocity 
  at $z=0$. We can easily extend the model to other redshifts
  by determining the effective e.o.s and the criterion for
  condensation as a function of redshift. We derive the time-dependent
  e.o.s. by integrating \Eq{eq-temp-evolve}, see \Sec{sec-eos}. 
  To obtain a time-dependent condensation criterion we determine 
  the position of the maximum in the density-temperature relation,
  see the solid line in \Fig{fig-Tn-delta-evolve}. To compensate for
  the discrepancy between $\Delta_{\rm max}$ and the actual overdensity
  for condensation we introduce a correction factor. To match the
  characteristic circular velocity obtained from the 
  simulations this factor also comprises the effect of the limited 
  numerical resolution. \Eq{eq-central-gas-density} can be now rewritten
	\begin{equation}
	  \left( \frac{v_{1/2}}{5.6 \: {\rm km \, s^{-1}} } \right)^2
	  =
	  \Delta_{\rm max}^\alpha \:
	  f_{\rm corr} \:
	  \frac{T_0}{10^4 \:{\rm K} } \:
	  \frac{\alpha+1}{\alpha} 
	  ,
	  \label{eq-central-gas-density-simple}
	\end{equation}
  where we have neglected the contribution of the outer halo boundary. To
  match the characteristic circular velocity at $z=0$ we choose $f_{\rm
  corr} = 2.5$. \Eq{eq-central-gas-density-simple} indicates that the
  characteristic circular velocity is essentially determined by the
  density-temperature relation at any time. All parameters on the r.h.s.,
  namely slope and normalisation of the e.o.s. and $\Delta_{\rm max}$, are
  derived from $T(\Delta,t)$. \Fig{fig-Tn-redshift-evolution} (upper row)
  shows how the characteristic circular velocity is composed of three
  parameters of the temperature-density relation. The upper left panel
  shows $T(\Delta)$ for several redshifts. The next three panels show the
  evolution of the parameters of $T(\Delta,t)$: The temperature $T_0$, the
  overdensity at the maximal temperature, $\Delta_{\rm max}$, and the
  slope ${\rm d} \log T(\Delta,t) / {\rm d} \log \Delta$  at $\Delta = 1$.
  The right panel shows the resulting $v_{1/2}$, which nicely matches the
  characteristic circular velocity obtained from the numerical
  simulations. In the range $ 3 \gtrsim z \geq 0$ the circular velocity is
  virtually constant. This is the result of the decreasing $T_0$ and the
  increasing $\Delta_{\rm max}$.
  \\

  The redshift evolution of the characteristic mass can be derived from
  characteristic circular and \Eq{eq-vmax-Mvir-rvir} and
  \Eq{eq-Mvir-Deltac}. Combining these equations and using $\langle \rho
  \rangle \propto (z+1)^3$ we find
	\begin{equation}
	  M_{1/2}
	  \propto
	  v_{1/2}^3 \:
	  \Delta_{\rm c}^{-1/2}(z) \:
	  (z+1)^{-3/2}
	  .
	  \label{eq-Mchar-vchar-z}
	\end{equation}  
  This is in agreement with the results found earlier \cite{hoeft:06}. More
  precisely, the dominating terms in $M_{1/2}$ are introduced by the redshift
  dependence in the definition of the virial radius, \Eq{eq-Mvir-Deltac}.

\subsection{Variations in the UV background}

\label{sec-vary-uvb}

\begin{figure*}
  \begin{center}
  \includegraphics[width=0.9\textwidth,angle=0]{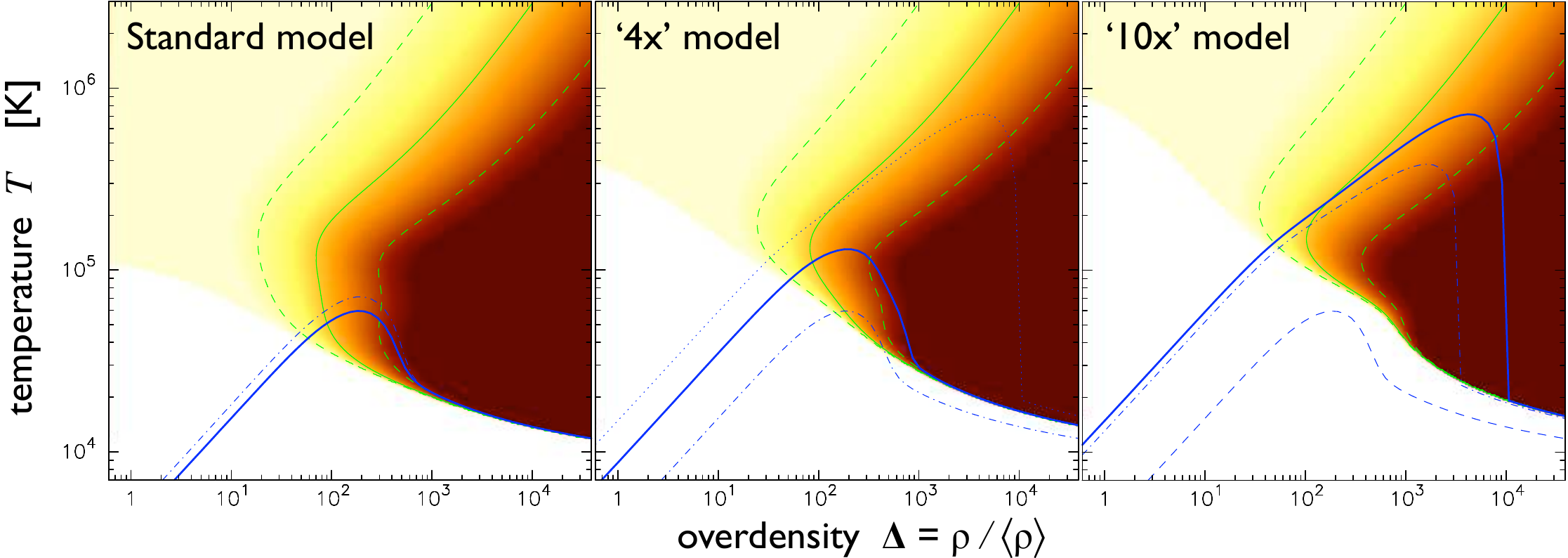}
  \end{center}
  \vspace{-0.6cm}
  \caption
  {
  Theoretical results for temperature distribution in the IGM as a result
  of different UVB heating models. {\it Left panel:} The thick solid line
  indicate the result for our standard model according to
  \cite{haardt:96}. Color coded is the cooling time. The blank area
  indicates the region where the gas is heated. Three contour lines are
  shown, they indicate from the left to right a cooling time 10x, 1x, and
  0.1x the Hubble time. The dashed blue line indicates a four times
  increased energy per ionising photon only for $z>2$. {\it Middle panel:}
  Results for four times increased energy per ionising photon for all redshifts.
  For comparison the dashed blue line indicates temperature distribution
  resulting from the standard model as shown in the left panel. {\it Right
  panel:} Results for ten times increased energy per ionising photon for all
  redshifts. For comparison also the eight times increased energy and the
  standard heating as shown in the left panel are indicated.
  }
  \label{fig-Tn-uv-models}
\end{figure*}

  As shown above the cumulative heating of the IGM by the UV background
  prevents the gas cooling in the centre of small mass haloes. It is
  therefore evident that the characteristic mass depends on the UV
  background and its evolution over cosmic times. We use our model worked
  out above to assess the impact of modifications of the UV background
  model on the characteristic circular velocity. We first consider
  changes of the UV flux density and secondly of the UV spectrum.

\paragraph*{UV flux density}

  The flux density of the UV background has only a small effect on the
  temperature of the IGM. The reason is that a higher flux reduces the
  fraction of neutral hydrogen in the tenuous IGM, hence, the higher flux
  finds fewer neutral atoms to ionise. As a result the heating rate is
  essentially independent from the flux density, when photo-ionisation
  dominates over collisional ionisation and as long as the spectrum
  remains unchanged. In this case the fraction of neutral hydrogen,
  $x_{\rm HI} = 1-x_{\rm HII}$, can be derived from the balance between
  photo-ionisation and recombination
	\begin{equation}
	  \Gamma_{\rm \gamma HI} \,
	  x_{\rm HI}
	  n_{\rm H}
	  =
	  \alpha_{\rm HII}(T) \,
	  x_{\rm HII}
	  n_{\rm H} \,
	  n_{\rm e}
	  .
	\end{equation} 
  The IGM is generally highly ionised, hence $x_{\rm HII}\sim 1$. In this
  case the electron density is proportional to the hydrogen density,
  $n_{\rm e} \sim n_{\rm H}$. As a result the heating due to hydrogen
  ionisation becomes
	\begin{equation}
	  {\cal H}_{\rm HI}
	  =
	  n_{\rm HI}
	  \epsilon_{\rm HI}
	  \sim
	  \alpha_{\rm HII}(T)
	  \frac{ \epsilon_{\rm HII} }
		   { \Gamma_{\rm \gamma HI} }
	  n_{\rm H}^2   
	  .  
	  \label{eq-low-density-heating}
	\end{equation}
  The heating therefore depends only on the ratio $\epsilon_{\rm
  HII}/\Gamma_{\rm \gamma HI}$, \ie only on the average energy input per
  ionisation event. In the limits made above the heat input is independent
  from the actual flux density. In accordance with this result we have
  found in Ref.\,\cite{hoeft:06} that the characteristic mass depends only
  little on the UV background flux density. Even very strong variations of
  the flux causes only a small shift of the characteristic mass, showing
  that the latter is sensitive only to the thermal state of the IGM which
  is little affected by the actual flux density.

\paragraph*{UV spectrum}

  Significantly more important for the thermal evolution of the IGM is the
  spectrum of the UV background. As a first guess one may expect that the
  final temperature of the IGM is directly proportional to the energy per
  ionising photon. However, since the recombination rate, $\alpha_{\rm
  HII}(T)$, decreases with increasing temperature the effect is not that
  strong. \Fig{fig-Tn-uv-models} shows the results for increased energies
  per ionising photon compared to the \cite{haardt:96} model. We have
  computed models with four times and with eight times increased energy per
  ionisation event. The higher average energy shifts the effective
  equation-of-state to higher temperatures. Also the shape of the
  temperature-density relation changes. The maximum in the relation shifts
  to higher densities. In particular, the maximum changes significantly
  between the four times and eight times increased model. The effective
  equation-of-state runs in the cooling regime virtually parallel to lines
  of constant cooling times. When $T_0 \gtrsim 1.5\times 10^4 \:{\rm K}$
  these cooling times are above the Hubble time. In those cases the cases
  the transition to thermal balance appears at very high overdensities.
  \\

  We use the models introduced above to estimate the impact of the UV
  spectrum on the characteristic circular velocity. To this end we compute
  the evolution of the characteristics of the temperature-density
  relation, see \Fig{fig-Tn-redshift-evolution}. For the four times
  increased energy per ionising photon we find that the characteristic
  circular velocity amounts to $\sim 60 \:{\rm km \, s^{-1}}$ for
  $z\lesssim 2$. For the eight times increased model the characteristic
  velocity get very high, since $\Delta_{\rm max}$ shifts to very high
  overdensities.


\section{Summary}
\label{sec-summary}

  We discuss high-resolution simulations including radiative cooling and
  photo-heating. We focus on the baryon content of dwarf galaxy-sized dark
  matter haloes. To this end, we have simulated the structure formation in two
  distinct cosmological environments and we found that the characteristic mass
  below which haloes are baryon deficient does not depend on the environment.
  Moreover, we found that in haloes in the mass range of $\sim 10^{9}$ to 
  $10^{10}\:\hMsun$ in general gas condensation stops at some redshift between
  reionisation and today. In our simulation those haloes are able to retain the
  already condensed gas, allowing further star formation. We have derived the
  characteristic mass in newly accreted matter, which is somewhat higher than
  the characteristic mass of the baryon fraction in the haloes. The former is
  more appropriate for semi-analytic modelling. 
  \\

  Tinker and Conroy \cite{tinker:09} found by a Halo Occupation Distribution
  analysis that the magnitude of galaxies which reside in haloes with a mass
  about the characteristic mass is $M_r \sim -10$. Moreover, they found
  that more luminous void galaxies, \ie $M_r > -10$, are
  preferably located close to the walls of the voids. Therefore, photo-heating
  would reduce the baryon content of the small-mass haloes populating the
  entire void volume, rendering these galaxies even more dark. Current surveys
  which cover large void volumes are not sensitive enough to detect the
  galaxies affected by photo-heating. 
  \\

  The failure of condensing gas can be traced back to the cooling time in the
  halo centre, which may exceed the Hubble time. We have set up a spherical
  model for the haloes gas density profile. Crucial for this model is the
  effective equation of state which we have derived by integrating the thermal
  history of the IGM using an approximation for the overdensity evolution as a
  function of the final overdensity We have justified this model by
  considering the density contrast evolution in the simulation. This spherical
  model allows us to derive the characteristic mass for different heating
  histories of the IGM. 
  \\

  We find that the mass at which galaxy formation seems to fade can be
  explained by a heating model which incorporates 6-8 times more energy per
  ionising photon than given in the Haardt \& Madau model \cite{haardt:96}.
  For redshift $z\gtrsim 3$ it is known from analyses of the Lyman$\alpha$
  forest that the temperature of the IGM is higher than obtained by a naive
  application of the Haardt \& Madau model. The derived temperature implies a
  3-4 times increased energy per ionising photon, or a more efficient heating
  due to non-equilibrium effects or radiative transfer effects. We found that
  if the average temperature of the IGM is today as high as $\gtrsim
  10^4\:{\rm K}$ dwarf galaxy formation would be suppressed at a mass scale
  consistent with that derived, \eg, by the conditional luminosity function.
  Therefore, the mass scale for suppression of dwarf galaxy formation may be
  considered as a measure for the temperature of gas in the surroundings of
  dwarf galaxies. The heat source could be the UV background or alternatively
  any feedback of the galaxy.
  \\

  \vspace{0.3cm}

  {\bf Acknowledgement:} The simulations discussed here have been performed at
  the LRZ Munich and the NIC J\"ulich. We would like to thank Gustavo Yepes
  and Volker Springel for many comments and discussions during the last years.
  Finally, we would like to thank our referees, in particular Michael Vogeley,
  who read the paper very carefully and provided many comments which improved
  this text.

\bibliography{void}{}
\bibliographystyle{plain}
 
\end{document}